\documentclass[twocolumn,trackchanges]{aastex631}

\newcommand{\be}{\begin{equation}}
\newcommand{\ee}{\end{equation}}
\usepackage{amsmath}
\usepackage{xcolor}
\usepackage{url}

\begin{document}

\title{\textit{XL-Calibur} polarimetry of Cyg X-1 further constrains the origin of its hard-state X-ray emission}

\correspondingauthor{*Ephraim Gau (ephraimgau@wustl.edu)}
\correspondingauthor{$^{**}$Kun Hu (hkun@wustl.edu)}
\correspondingauthor{$^{\dag}$Mózsi Kiss (mozsi@kth.se)}
\correspondingauthor{$^{\ddag}$Sean Spooner (Sean.Spooner@unh.edu)}

\author[0009-0001-3135-0754]{Hisamitsu Awaki} \affiliation{Graduate School of Science and Engineering, 2-5, Bunkyo-cho, Matsuyama, Ehime 790-8577, Japan} 
\author[0000-0003-4433-1365]{Matthew G. Baring} \affiliation{Department of Physics and Astronomy -- MS 108, Rice University, 6100 Main Street, Houston, Texas 77251-1892, USA}
\author{Richard Bose} \affiliation{Department of Physics, McDonnell Center for the Space Sciences and Center for Quantum Sensors, Washington University in St. Louis, 1 Brookings Dr, Saint Louis, MO 63130, USA} 
\author[0009-0009-3051-6570]{Jacob Casey} \affiliation{Department of Physics and Astronomy, and Space Science Center, University of New Hampshire, 8 College Rd, Durham, NH 03824, USA}
\author[0009-0002-2488-5272]{Sohee Chun} \affiliation{Department of Physics, McDonnell Center for the Space Sciences and Center for Quantum Sensors, Washington University in St. Louis, 1 Brookings Dr, Saint Louis, MO 63130, USA}
\author{Adrika Dasgupta} \affiliation{Department of Physics and Astronomy, and Space Science Center, University of New Hampshire, 8 College Rd, Durham, NH 03824, USA}
\author[0000-0001-6358-5147]{Pavel Galchenko} \affiliation{NASA Wallops Flight Facility, Fulton St, Wallops Island, VA 23337, USA}
\author[0000-0002-5250-2710]{Ephraim Gau*} \affiliation{Department of Physics, McDonnell Center for the Space Sciences and Center for Quantum Sensors, Washington University in St. Louis, 1 Brookings Dr, Saint Louis, MO 63130, USA}
\author{Kazuho Goya} \affiliation{Graduate School of Advanced Science and Engineering, Hiroshima University, 1-3-1 Kagamiyama, Higashi-Hiroshima, Hiroshima 739-8526, Japan}
\author[0000-0002-0987-0278]{Tomohiro Hakamata} \affiliation{Department of Earth and Space Science, Osaka University, 1-1 Machikaneyama-cho, Toyonaka, Osaka 560-0043, Japan} 
\author[0000-0001-6665-2499]{Takayuki Hayashi} \affiliation{NASA Goddard Space Flight Center, Greenbelt, MD 20771, USA} \affiliation{University of Maryland, Baltimore County, 1000 Hilltop Circle, Baltimore, MD 21250, USA} 
\author{Scott Heatwole} \affiliation{NASA Wallops Flight Facility, Fulton St, Wallops Island, VA 23337, USA}
\author[0000-0002-9705-7948]{Kun Hu$^{**}$} \affiliation{Department of Physics, McDonnell Center for the Space Sciences and Center for Quantum Sensors, Washington University in St. Louis, 1 Brookings Dr, Saint Louis, MO 63130, USA} 
\author{Daiki Ishi} \affiliation{Japan Aerospace Exploration Agency, Institute of Space and Astronautical Science, 3-1-1 Yoshino-dai, Chuo-ku, Sagamihara, Kanagawa 252-5210, Japan} 
\author[0000-0003-4509-3493]{Manabu Ishida} \affiliation{Japan Aerospace Exploration Agency, Institute of Space and Astronautical Science, 3-1-1 Yoshino-dai, Chuo-ku, Sagamihara, Kanagawa 252-5210, Japan} 
\author[0000-0001-7477-0380]{Fabian Kislat} \affiliation{Department of Physics and Astronomy, and Space Science Center, University of New Hampshire, 8 College Rd, Durham, NH 03824, USA}
\author[0000-0001-5191-9306]{M\'ozsi Kiss$^{\dag}$} \affiliation{KTH Royal Institute of Technology, Department of Physics, 106 91 Stockholm, Sweden} \affiliation{The Oskar Klein Centre for Cosmoparticle Physics, AlbaNova University Center, 106 91 Stockholm, Sweden} 
\author{Kassi Klepper} \affiliation{KTH Royal Institute of Technology, Department of Physics, 106 91 Stockholm, Sweden} \affiliation{The Oskar Klein Centre for Cosmoparticle Physics, AlbaNova University Center, 106 91 Stockholm, Sweden} 
\author[0000-0002-1084-6507]{Henric Krawczynski} \affiliation{Department of Physics, McDonnell Center for the Space Sciences and Center for Quantum Sensors, Washington University in St. Louis, 1 Brookings Dr, Saint Louis, MO 63130, USA}
\author[0009-0006-9014-2716]{Haruki Kuramoto} \affiliation{Department of Earth and Space Science, Osaka University, 1-1 Machikaneyama-cho, Toyonaka, Osaka 560-0043, Japan}
\author[0000-0002-5202-1642]{Lindsey Lisalda} \affiliation{Department of Physics, McDonnell Center for the Space Sciences and Center for Quantum Sensors, Washington University in St. Louis, 1 Brookings Dr, Saint Louis, MO 63130, USA} 
\author[0000-0002-9099-5755]{Yoshitomo Maeda} \affiliation{Japan Aerospace Exploration Agency, Institute of Space and Astronautical Science, 3-1-1 Yoshino-dai, Chuo-ku, Sagamihara, Kanagawa 252-5210, Japan} 
\author{Hironori Matsumoto} \affiliation{Department of Earth and Space Science, Osaka University, 1-1 Machikaneyama-cho, Toyonaka, Osaka 560-0043, Japan} \affiliation{Forefront Research Center, Graduate School of Science, Osaka University, Japan}
\author[0009-0005-0818-7484]{Shravan Vengalil Menon} \affiliation{Department of Physics, McDonnell Center for the Space Sciences and Center for Quantum Sensors, Washington University in St. Louis, 1 Brookings Dr, Saint Louis, MO 63130, USA}
\author[0009-0009-2127-8830]{Aiko Miyamoto} \affiliation{Department of Earth and Space Science, Osaka University, 1-1 Machikaneyama-cho, Toyonaka, Osaka 560-0043, Japan}
\author[0009-0008-4216-064X]{Asca Miyamoto} \affiliation{Department of Physics, Tokyo Metropolitan University, 1-1 Minami-Osawa, Hachioji, Tokyo 192-0397, Japan}
\author{Kaito Murakami} \affiliation{Department of Earth and Space Science, Osaka University, 1-1 Machikaneyama-cho, Toyonaka, Osaka 560-0043, Japan} 
\author[0000-0002-6054-3432]{Takashi Okajima} \affiliation{NASA Goddard Space Flight Center, Greenbelt, MD 20771, USA} 
\author[0000-0001-7011-7229]{Mark Pearce} \affiliation{KTH Royal Institute of Technology, Department of Physics, 106 91 Stockholm, Sweden} \affiliation{The Oskar Klein Centre for Cosmoparticle Physics, AlbaNova University Center, 106 91 Stockholm, Sweden} 
\author[0000-0002-1452-4142]{Brian Rauch} \affiliation{Department of Physics, McDonnell Center for the Space Sciences and Center for Quantum Sensors, Washington University in St. Louis, 1 Brookings Dr, Saint Louis, MO 63130, USA} 
\author[0000-0001-5256-0278]{Nicole Rodriguez Cavero} \affiliation{Department of Physics, McDonnell Center for the Space Sciences and Center for Quantum Sensors, Washington University in St. Louis, 1 Brookings Dr, Saint Louis, MO 63130, USA} 
\author{Kentaro Shirahama} \affiliation{Department of Earth and Space Science, Osaka University, 1-1 Machikaneyama-cho, Toyonaka, Osaka 560-0043, Japan} 
\author[0000-0003-0710-8893]{Sean Spooner$^{\ddag}$} \affiliation{Department of Physics and Astronomy, and Space Science Center, University of New Hampshire, 8 College Rd, Durham, NH 03824, USA} 
\author[0000-0001-6314-5897]{Hiromitsu Takahashi} \affiliation{Graduate School of Advanced Science and Engineering, Hiroshima University, 1-3-1 Kagamiyama, Higashi-Hiroshima, Hiroshima 739-8526, Japan} 
\author[0009-0004-7101-4503]{Keisuke Tamura} \affiliation{NASA Goddard Space Flight Center, Greenbelt, MD 20771, USA} \affiliation{University of Maryland, Baltimore County, 1000 Hilltop Circle, Baltimore, MD 21250, USA} 
\author[0000-0002-7962-4136]{Yuusuke Uchida} \affiliation{Tokyo University of Science, 2641 Yamazaki, Noda, Chiba 278-8510, Japan}
\author[0000-0003-4434-1766]{Kasun Wimalasena} \affiliation{Department of Physics and Astronomy, and Space Science Center, University of New Hampshire, 8 College Rd, Durham, NH 03824, USA}
\author{Masato Yokota} \affiliation{Graduate School of Advanced Science and Engineering, Hiroshima University, 1-3-1 Kagamiyama, Higashi-Hiroshima, Hiroshima 739-8526, Japan}
\author[0009-0005-0819-0819]{Marina Yoshimoto} \affiliation{Department of Earth and Space Science, Osaka University, 1-1 Machikaneyama-cho, Toyonaka, Osaka 560-0043, Japan}



\begin{abstract}
The balloon-borne hard X-ray polarimetry 
mission  {\it XL-Calibur} observed the Black Hole X-ray Binary (BHXRB) Cygnus X-1 (Cyg X-1) during its nearly six-day Long Duration Balloon (LDB) flight from Sweden to Canada in July 2024. 
The {\it XL-Calibur} observations allowed us to derive the most precise constraints to date of the Polarization Degree (PD) and Polarization Angle (PA) of the hard X-ray emission from a BHXRB.
{\it XL-Calibur} observed Cyg X-1 in the hard state and measured a $\sim$\,19\,--\,64\,keV PD of ($5.0^{+2.7}_{-3.0}$)\% 
at a PA of $-28^{\circ}\pm 17^{\circ}$, with
an 8.7\% chance probability 
of detecting larger PDs than the one observed, given an unpolarized signal.
The {\it XL-Calibur} results 
are thus comparable to the 2\,--\,8\,keV PD and PA found by {\it IXPE}, 
with a similar agreement between the hard X-ray PA and the radio jet direction.
We also discuss the implications of our polarization measurements in the context of models describing the origin of the broadband X-ray and $\gamma$-ray emission, to which {\it XL-Calibur} provides independent constraints on any proposed emission modeling. 
\end{abstract}

\keywords{(1) High Energy Astrophysics, (2) X-ray Polarization, (3) Balloon-borne Mission, (4) Stellar-mass Black Holes, (5) Black Hole X-ray Binaries, (6) Black Hole Accretion, (7) Black Hole Corona}



\section{Introduction}
\label{s:introduction}

The polarization of the 2\,--\,8\,keV X-ray emission 
from a variety of galactic and extragalactic sources,
including BHXRBs,
accretion-powered neutron stars,
magnetars, pulsar wind nebulae, 
supernova remnants, and 
Active Galactic Nuclei (AGNs)
\citep[e.g.,][]{2022Sci...378..650K,2022NatAs...6.1433D,2022Sci...378..646T, 2022Natur.612..658X, 2022ApJ...938...40V,2022Natur.611..677L, 2022ApJ...935..116E}
has been measured by the {\it Imaging X-ray Polarimetry Explorer} ({\it IXPE}) \citep{ixpe}. 
Most relevant to our study, {\it IXPE} has observed BHXRBs in several states, as covered in the recent review papers of \cite{2024mbhe.confE..39C} and \cite{2024Galax..12...54D}.

The {\it soft-state} emission from BHXRBs is dominated by the diluted multi-temperature blackbody emission from a geometrically thin, optically thick accretion disk. 
Measurements of this emission constrain the inclination of the accretion disk and the black hole spin (that is, the angle between the black hole spin axis and the line of sight to the observer) \citep{2009ApJ...691..847L, 2009ApJ...701.1175S}.
The {\it hard-state} emission includes the Comptonized (upscattered) emission from the coronal plasma, as well as coronal emission reflected by the disk
\citep{2010ApJ...712..908S, 2019ApJ...875..148Z, 2022ApJ...934....4K,2023ApJ...949L..10P}. 
One such BHXRB, Cyg X-1, comprises a (21.2$^{+2.2}_{-2.3}$)\,$M_{\odot}$ black hole, along with a ($40.6^{+7.7}_{-7.1})\,M_{\odot}$ companion, located at a distance of ($2.22^{+0.18}_{-0.17})\,$kpc from Earth \citep{2021Sci...371.1046M}.
It is found in the hard state about two-thirds of the time \citep{2006A&A...447..245W, 2013A&A...554A..88G}. 

The {\it IXPE} hard-state observations of Cyg X-1 revealed a 2\,--\,8\,keV PD of (4.01$\pm$0.20)\%, surprisingly high when compared to the predictions of laterally-extended corona models \citep[e.g.,][]{2010ApJ...712..908S,2022ApJ...934....4K}, 
as well as a PA of $-20^{\circ}.7 \pm 1^{\circ}.4$ (with positive angles denoting directions East of Celestial North), which is aligned with the axis of its radio jet.
The measurements indicated an increase of the PD with energy, going from (3.6$\pm$0.4)\% to (5.7$\pm$0.9)\% between 2\,keV to 8\,keV \citep{2022Sci...378..650K}.
{\it IXPE} revealed a similar alignment of the coronal emission polarization with the radio jets 
of the BHXRBs Swift\,J1727.8$-$1613 \citep{2023ApJ...958L..16V, 2024ApJ...968...76I, 2024ApJ...971L...9W} and GX\,339$-$4 \citep{2024arXiv240806856M}, 
as well as the AGNs NGC\,4151 \citep{1982ApJ...262...61J, 1998ApJ...496..196U, 2023MNRAS.523.4468G, 2024A&A...691A..29G} and IC\,4329A \citep{1987MNRAS.228..521U, 2023MNRAS.525.5437I}.
This alignment thus seems to be a general feature of the coronal emission from accretion disks across a broad range of black hole masses \citep{2024ApJ...974..101S}.
{\it IXPE} soft state observations of Cyg X-1 
showed the PA still being aligned with the radio jet but  with a somewhat lower 2\,--\,8\,keV PD of (1.99$\pm$0.13)\% \citep{2024ApJ...969L..30S}.

The hard state polarization results of BHXRBs from {\it IXPE} can be interpreted within different theoretical frameworks.
In one family of models,  
the power-law X-rays originate within a horizontally-extended corona Comptonizing 
lower-energy synchrotron emission or standard/truncated accretion disk emission in a hot $T_{\rm C}\sim\,$100\,keV plasma
\citep[e.g.][and references therein]{1985A&A...143..374S,1993A&A...275..337P,2010ApJ...712..908S, 2022ApJ...934....4K}.
The high 2\,--\,8\,keV PDs measured by {\it IXPE} in \citet{2022Sci...378..650K} could be explained by the models analyzed in that study, 
but only if the inner disk inclination is $\ge$\,$45^{\circ}$, substantially higher than the $(27^{\circ}.5^{+0^{\circ}.8}_{-0^{\circ}.6})$ inclination of the overall binary system \citep{2021Sci...371.1046M}. 
However, the inclination constraints on those models can be softened by invoking Comptonization in a mildly-relativistically outflowing corona \citep{Beloborodov_1999,2023ApJ...949L..10P}.
Moreover, Particle In Cell (PIC) and resistive General Relativistic Magnetohydrodynamic (rGRMHD) simulations, in combination with general considerations, reveal an alternative scenario in which the Comptonization is effected by (likely cold) plasmons, generated by magnetic reconnection,
plasma turbulence, or both
\citep{2017ApJ...850..141B,2025ApJ...979..199S}.
In this context, ``cold'' refers to the fact that the plasma bulk motion (which is mildly relativistic, at most), rather than the random motion of the plasma particles, dominates the photon energization.
Alternative models invoke scattering off cold plasma outflows \citep{1987ApJ...322..650B, 2024MNRAS.528L.157D}
or synchrotron and inverse Compton emission from the jet or jet walls \citep{2024Ap&SS.369...68M}.

With an energy range of $\sim$\,19\,--\,64\,keV, the balloon-borne hard X-ray spectropolarimeter {\it XL-Calibur} complements the 2\,--\,8\,keV range of {\it IXPE}.
{\it XL-Calibur} flew on an LDB flight from Esrange, Kiruna, Sweden to near Kugluktuk, Nunavut, Canada for a roughly weeklong flight, observing Cyg X-1 (in its low/hard state) and the Crab Pulsar/Pulsar Wind Nebula. 
This paper presents the results from the Cyg X-1 observations. 
The results from the Crab observations, as well as additional details about the {\it XL-Calibur} flight and data analysis, can be found in \citet{2025MNRAS.tmpL..25A}. 

One important advantage of {\it XL-Calibur} is that the blackbody emission from the inner accretion disk is not thought to significantly contribute to the detected signal in the hard X-ray energy range, whereas such a component may still be sampled within the lower {\it IXPE} bandpass
\citep{1999ASPC..161...64G}.
Furthermore, depending on the model, direct coronal emission and coronal emission reprocessed and reflected off the accretion disk may contribute differently in the $\sim$\,19\,--\,64\,keV band than in the 2\,--\,8\,keV band. 

Three earlier experiments had reported Cyg X-1 polarization results in the hard X-ray and $\gamma$-ray bands. 
The balloon-borne X-ray polarimetry experiment {\it PoGO+} measured an upper limit of $<$8.6\% (at the 90\% confidence level) and a point measurement of ($0.0^{+5.6}_{-0.0}$)\% for the 19\,--\,181\,keV PD \citep{2018NatAs...2..652C}.
At even higher, $>$100\,keV energies, {\it AstroSat}, {\it INTEGRAL}/IBIS, and {\it INTEGRAL}/SPI observations revealed evidence for polarized emission, likely resulting from the Cyg X-1 jet---the PD they detected is much higher, and the PA is significantly different from those at soft X-ray energies  \citep{2024ApJ...960L...2C,2011Sci...332..438L, 2012ApJ...761...27J, 2015ApJ...807...17R}.
The data from {\it XL-Calibur} thus bridges the gap between the soft X-ray and $\gamma$-ray bands with the highest precision polarization measurements of Cyg X-1 in the hard X-rays to date, 
allowing for an independent constraint on any models invoked to explain the observed emission across the high-energy bandpass.

The rest of the paper is organized in the following manner. 
We give an overview of the {\it XL-Calibur} mission and data analysis methods in Section \ref{s:xlcalibur}.
The observations of Cyg X-1 conducted during the 2024 LDB flight are described in Section \ref{s:observation}. 
We then present our Cyg X-1 polarization results in Section \ref{s:results}, 
before discussing the implications of these results for the origin of the hard X-ray emission, especially in the context of broadband high-energy polarization results, in Section \ref{s:discussion}.


\section{The {\it XL-Calibur} Mission}
\label{s:xlcalibur}

\subsection{Telescope design}
\label{s:instrument}

{\it XL-Calibur} \citep{2021APh...12602529A} uses a 12\,m optical bench (truss), composed of carbon-fiber tubes and aluminum joints.
The truss supports 
both the X-ray mirror (almost identical to the mirror onboard {\it Hitomi} described by \citet{2014ApOpt..53.7664A}) on the front end 
and the rotating scattering polarimeter \citep{2021APh...12602529A} 
enclosed within a $\sim$\,3.5\,cm thick anticoincidence Bi$_{4}$Ge$_{3}$O$_{12}$ (BGO) shield on the rear end \citep{2023NIMPA104867975I}. 
The Wallops Arc Second Pointer (WASP) \citep{doi:10.2514/6.2017-3609, 2022JGCD...45.1365G}, pointing the telescope towards each astrophysical source of interest, is mounted on the gondola connecting the truss to the balloon. During the 2024 flight, it achieved a pointing precision of $<$\,10\,arcsec for $>$\,96\% of the observation time. 

The X-ray mirror is made of 213 concentric Al shells coated with platinum-carbon multilayers. The mirror  
achieves a Point Spread Function (PSF) with a half-power diameter of 2\,arcmin 
and an effective area of $\sim$\,300\,cm$^2$ at 20\,keV and $\sim$\,50\,cm$^2$ at 60\,keV \citep{2023SPIE12679E..1BK}.

The scattering polarimeter is composed of an 80\,mm long, 12\,mm diameter, low-atomic-number beryllium (Be) rod 
surrounded by sixteen 0.8\,mm thick, 20$\times$20\,mm$^2$ footprint, 64-pixel, high-atomic-number ($Z\,\sim\,\,$50) Cadmium Zinc Telluride (CZT) detectors.
An identical 17$^{\rm th}$ (imaging) CZT detector is positioned behind the scattering rod to capture unscattered 
X-rays. 
The entire shield and polarimeter assembly is rotated at a constant speed of $\sim$\,2 revolutions-per-minute throughout the flight, 
in order to minimize systematic biases due to azimuthal variations of efficiency, threshold, and angular coverage between different pixels.

Photons preferentially scatter in a direction perpendicular to their electric vector position angle (the EVPA, same as the PA) according to the Klein-Nishina distribution, 
leading to an azimuthal scattering angle distribution of
\begin{eqnarray*}
dN/d\psi\, \propto \, \frac{N}{2\pi} (1 + p_0 \, \mu \, \cos(2(\psi - \psi_0 - \pi/2)),
\end{eqnarray*}
where $p_0$ is the true PD of the incident beam, $\psi_0$ is its PA, 
and $\mu$ is the instrument- and observation-specific modulation response ($\sim$\,0.43 for {\it XL-Calibur} observing Cyg X-1 in the hard state)
for a 100\% linearly-polarized beam
\citep{2024APh...15802944A}.\footnote{As \citet{2025MNRAS.tmpL..25A} describes, the overall modulation factor is likely only known to a (systematic) precision of 0.002. 
However, this is negligible in comparison with the statistical error/precision that will be stated for the main results of this paper.}

The atmospheric absorption of lower-energy photons limits the {\it XL-Calibur} energy range to photons $\gtrsim$\,15\,keV as measured in the CZTs.
We used pixel-dependent energy thresholds to remove lower-energy noise,
with the median pixel threshold being $\sim$\,15.4\,keV with a standard deviation of $\sim$\,4.7\,keV.
The signal flux drops off rapidly at higher energies and goes below the background $\sim$\,60\,keV as measured in the CZTs.
Taken together, these effects limit the effective signal energy range to 15\,--\,60\,keV as measured in the CZTs after scattering, which corresponds to the range of $\sim$\,19\,--\,64\,keV for incident photons, as based on simulations \citep{2024APh...15802944A}
wherein physical effects such as Compton scattering and the energy resolution (3.8\,keV at 20\,keV and 11.7\,keV at 64\,keV, as detailed in \citet{2021APh...12602529A}) have been accounted for. 

\subsection{Data analysis}
\label{s:analysis}

The data analysis uses events that 
triggered only one CZT pixel and that
passed the cuts suppressing noisy channels.
Additionally, only events without a BGO shield veto, and of energies between 15\,keV and 60\,keV as recorded in the CZT detectors, corresponding to $\sim$\,19\,--\,64\,keV for the incident photons (see \citet{2025MNRAS.tmpL..25A} for further discussion), are used.

For each event, an azimuthal scattering angle is calculated by assuming that the photon scattered off an offset-corrected scattering location into the CZT channel where it was detected, where the offset used is given by the background-subtracted mean position of counts in the imaging CZT detector for each day (see \citet{2024APh...15802944A} for a discussion of the necessity and benefit of such a correction). 
The azimuthal scattering angle is then converted into the Stokes parameters \citep{2015APh....68...45K} and 
referenced to Celestial North by accounting for 
the rotation of the polarimeter, 
the bank of the telescope (as calculated by the WASP star trackers and inertial navigation unit), 
and the 90$^{\circ}$ angular difference between the preferred scattering direction from the actual PA.  
We then performed an exposure-weighted background subtraction to obtain the Stokes parameters of the Cyg X-1 signal. 
 
Following the approach in \cite{2025MNRAS.tmpL..25A}, initial point-estimates for the PD\footnote{
The PD equation is sometimes seen with an extra factor of 2 in the numerator, as in \citet{2015APh....68...45K} and \citet{Kiss2024}.
However, as in the convention used by \citet{Kislat2024}, we have here absorbed that factor into the definition of the Stokes parameters themselves.
} and PA are defined by
\begin{eqnarray*}
PD &=& \frac{1}{\mu (I_{on} - I_{off})} \sqrt{(Q_{on} - Q_{off})^2 + (U_{on} - U_{off})^2}\\
PA& =& \frac{1}{2}\,\arctan\left(\frac{U_{on} - U_{off}}{Q_{on} - Q_{off}}\right).
\end{eqnarray*}

In the case of Cyg X-1, the PD magnitude is comparable to the minimum detectable polarization at 99\% confidence level ({\it MDP}$_{99}$, the smallest PD that could be detected at this confidence level),
a point at which the bias resulting from the positive-definiteness of the PD will affect the result \citep{2025MNRAS.tmpL..25A}.
We thus use a Bayesian analysis for 
determining the probability density distribution of the posterior.
A prior of $1/\sqrt{(Q/I)^2 + (U/I)^2}$ 
is used, which presupposes that all PD and PA values are equally likely \citep{2012A&A...538A..65Q, 2017NatSR...7.7816C, 2020ApJ...891...70A, Kiss2024}.
Finally, the 1$\sigma$ error intervals of the PD (positive-definite) and of the PA are found through marginalizing the posterior over the other variable
and 
calculating the values encompassing the densest 68.27\% of the total probability.
Even in the cases when a measurement is below the {\it MDP}$_{99}$, useful constraints on the PA can still be derived from such Bayesian analyses \citep{2025arXiv250404775L}.


\section{The Cyg X-1 Observations and Lightcurves}
\label{s:observation}

\begin{figure}
  \centering
  \includegraphics[width=0.475\textwidth]{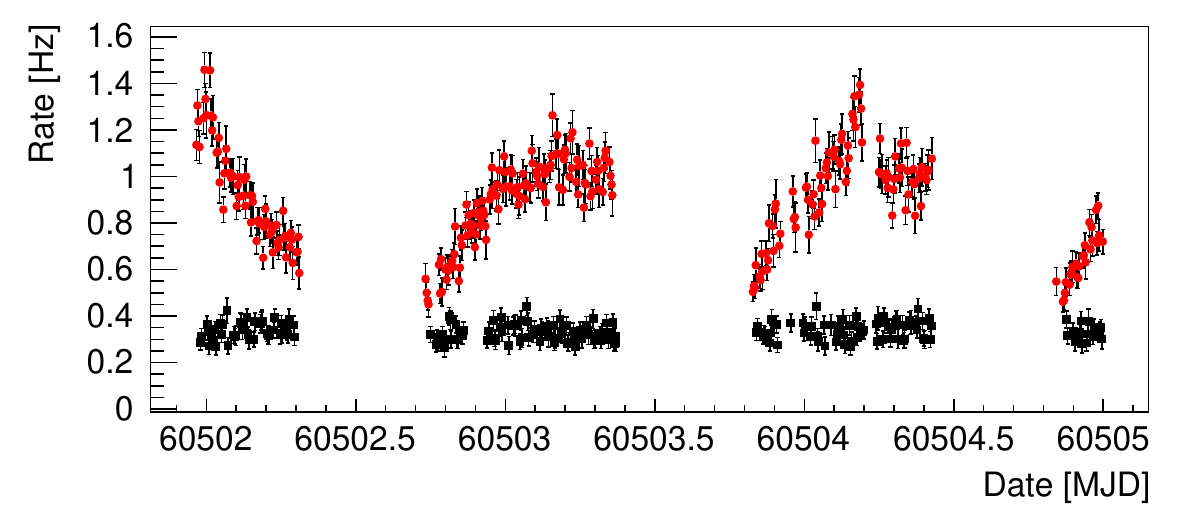}
  \caption{
  On-source (red circles) and off-source/background (black squares) events, of energies between $\sim$\,19\,--\,64\,keV and arriving in the polarimeter detectors, passing the analysis cuts described in Sect.\,\ref{s:analysis}. 
  Time bins of duration no longer than 300\,s are used.
  The daily gaps between Cyg X-1 observations were used for observing the Crab \citep{2025MNRAS.tmpL..25A}.
  Smaller variations within each day (for example, the drop in the third observation) are generally correlated with changes in the elevation of the source.
  }
  \label{fig:lightcurve}
\end{figure}

{\it XL-Calibur} was launched from the Esrange Space Center in Sweden on 2024-07-09 at 03:04 UTC and landed near Kugluktuk in Canada on 2024-07-14 at 23:19 UTC. 
About forty-three hours 
were spent observing Cyg X-1 across four days, at elevations between $\sim$\,25$^{\circ}$ and $\sim$\,55$^{\circ}$. 
During these observations, the telescope performed an ON/OFF cross-like nodding pattern centered on the source, with ON observations of 18\,minutes each, interspersed with OFF observations (at an offset of 1$^{\circ}$ from the source) of 12\,minutes each.

The rates obtained in the polarimeter during the ON and OFF observations are shown in the lightcurves of Fig.\,\ref{fig:lightcurve}, which displays a clear detection of a Cyg X-1 signal rate of up to $\sim$\,1.0\,Hz above a stable background rate of $\sim$\,0.35\,Hz.
The diurnal variations of the signal rate generally stem from the elevation-dependent atmospheric column density, 
which results in varying levels of detected flux as the source transits the sky each day. 

Across the four days of Cyg X-1 observations, a total number of $\sim$\,78,500 events were detected in the sixteen circumjacent (polarimeter) CZT detectors.
Out of these 78,500 events, approximately 62,400 arrive during on-source observations of Cyg X-1, and approximately 16,100 occur during off-source observations (purely background events).

\begin{figure}
  \centering
  \includegraphics[width=0.45\textwidth]{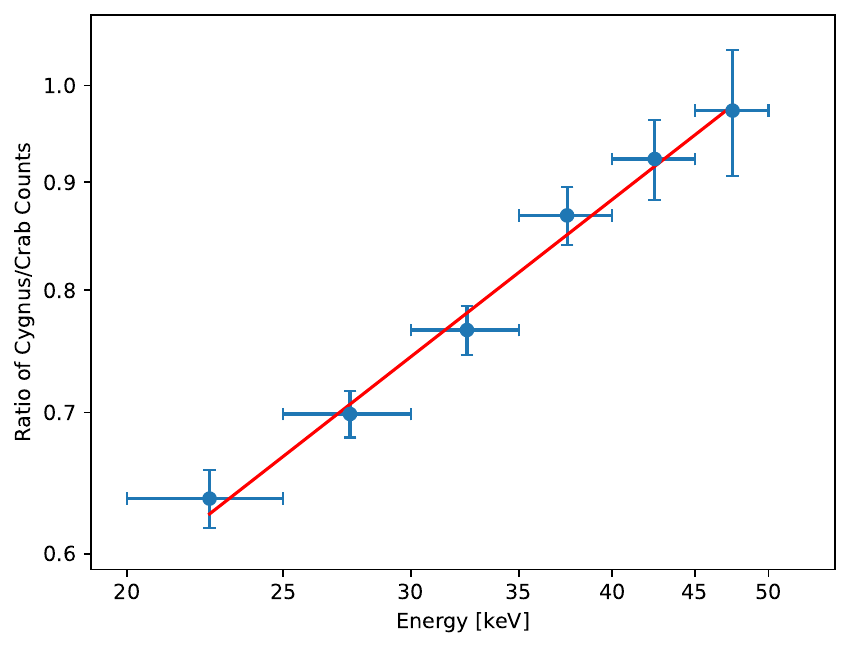}
  \caption{
  Ratio (blue data points) of the Cyg X-1 to the Crab background-subtracted signal count detection rates as a function of energy, with both axes in logarithmic scale.
  The counts are between 20\,keV to 50\,keV, as measured in only the polarimeter CZT detectors, and are binned in intervals of 5\,keV.
  The Cyg X-1 energy spectrum is harder than the Crab spectrum by  $\Delta\Gamma\,\approx\,0.60\pm0.03$.
  The red line gives the best linear fit to this ratio spectrum.
  The uncertainty on the $\Delta\Gamma$ is given by standard error propagation, taking into account the uncertainty arising both from the division of the two spectra as well as from the determination of the linear fit slope. 
  }
  \label{fig:cygnusCrabRatio}
\end{figure}

Fig.\,\ref{fig:cygnusCrabRatio} presents the ratio of the Cyg X-1 count-rate spectrum divided by the corresponding Crab spectrum, with both measured by {\it XL-Calibur} during the same balloon flight with a similar column density distribution (an average line-of-sight column density of 5.5 g/cm$^2$ for the Crab and 6.0 g/cm$^2$ for Cyg X-1). 
This ratio of energy spectra follows a $E^{\Delta \Gamma}$ power law, with $\Delta\Gamma\, \approx \,0.60\pm0.03$. 
The time-averaged energy spectrum of the Crab has a photon index of 
$\Gamma_{\rm Crab} = 2.10\pm0.02$, as described by \citet{2015ApJ...801...66M} for the 1\,--\,100\,keV bandpass (co-spectral with {\it XL-Calibur}), 
with $dN/dE \propto E^{-\Gamma_{\rm Crab}}$.\footnote{We have examined the flux data provided by {\it SWIFT}/BAT \citep{2013ApJS..209...14K}, and compared the flux recorded by that instrument (operating in the 15\,--\,150\,keV range) during the days that {\it XL-Calibur} observed the Crab, to the flux that it recorded during the time frames coincident with the ones used for the measurements of \citet{2015ApJ...801...66M}. In both cases, the flux is between 0.20 and 0.26 counts/cm$^2$/s.}
Thus, we estimate a Cyg X-1 photon index of $\Gamma_{\rm Cyg\,X-1}\, \approx \,\Gamma_{\rm Crab}-\Delta\Gamma$ $\approx\,1.50\pm0.04$ during our observations, consistent with Cyg X-1 being in the hard state \citep{2006A&A...447..245W}. 
A detailed spectral and spectropolarimetric analysis using complete instrument response matrices for {\it XL-Calibur}, in combination with contemporaneous {\it NICER} and {\it NuSTAR} observations, will be published in a forthcoming paper, which will permit a precise determination of the Cyg X-1 spectral index.


\section{Cyg X-1 Hard X-ray Polarimetric Results}
\label{s:results}

\begin{figure}
  \centering
  \includegraphics[width=0.475\textwidth]{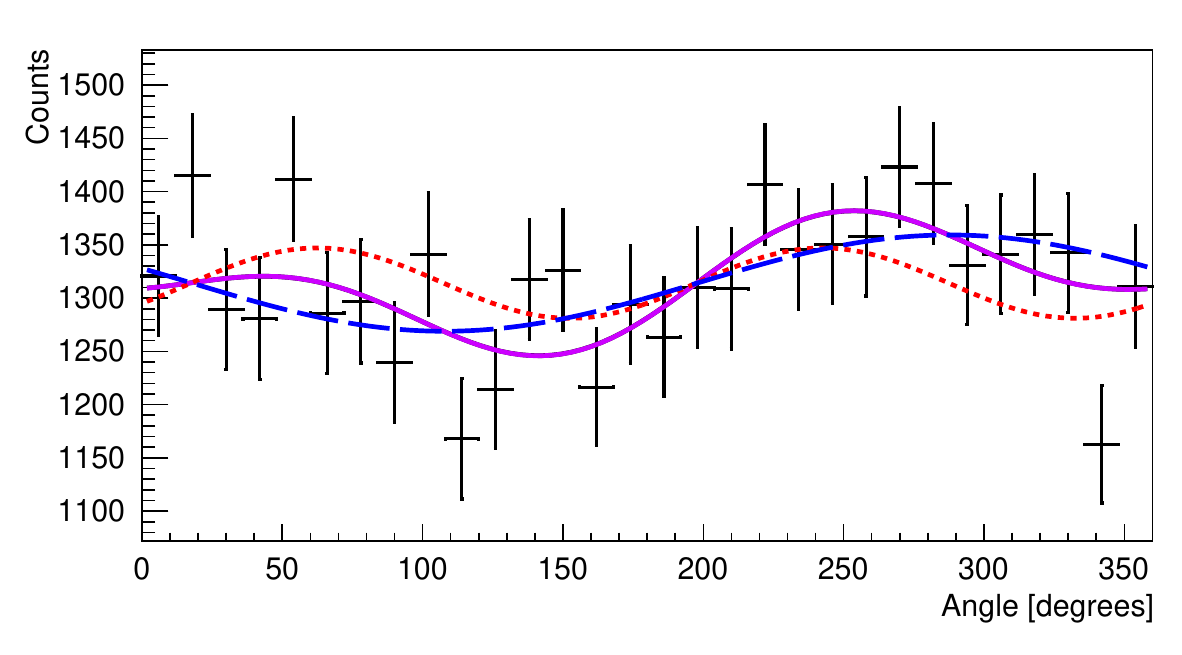}
  \caption{Azimuthal scattering angle distribution (black data points) of the $\sim$\,19\,--\,64\,keV background-subtracted Cyg X-1 signal arriving in the polarimeter detectors. The overall best-fit (purple solid line) is shown, as well as the 180$^{\circ}$ component (red dotted line) and 360$^{\circ}$ component (blue dashed line) contributions to the overall signal. 
  }
  \label{fig:modulation}
\end{figure}

Fig.\,\ref{fig:modulation} displays the background-subtracted azimuthal count distribution, fitted with a component of periodicity 180$^{\circ}$ as well as a component of periodicity 360$^{\circ}$. 
The former reflects the polarization of the signal; 
the latter mainly arises from an offset of the telescope's observing axis from the nominal scattering location at the center of the beryllium rod \citep{2025MNRAS.tmpL..25A}.\footnote{An expected azimuthal variation due to known asymmetries in the mirror PSF will also contribute.} 

\begin{table*}
    \centering
    \hspace*{-2cm}
    \begin{tabular}{cccccc}
        \hline
        Energy (keV) & $Q/I$ & $U/I$ & PD ($\%$) & PA($^{\circ}$) & $MDP_{99}$ ($\%$) \\
        \hline
        \hline
        $\sim$\,19\,--\,64 & $0.033 \, \pm \, 0.026$  & $-0.046 \, \pm \, 0.026$ & $5.0^{+2.7}_{-3.0}$ & $-28 \pm 17$ & 7.8 \\
        $\sim$\,19\,--\,35 & $0.016 \, \pm \, 0.035$ & $-0.028 \, \pm \, 0.035$ & $0.2^{+4.4}_{-0.2}$ & $-31 \pm 40.$  & 10. \\
        $\sim$\,35\,--\,64 & $0.051 \, \pm \, 0.038$ & $-0.066 \, \pm \, 0.038$ & $7.2^{+4.0}_{-4.4}$ & $-26 \pm 17$  & 12 \\
        \hline
    \end{tabular}
    \caption{
    The background-subtracted, normalized Stokes $Q$ and Stokes $U$ results, the marginalised PD and PA values from the Bayesian analysis, and the $MDP_{99}$ values for the {\it XL-Calibur} observation. 
    The two energy subdivisions are chosen so as to have approximately equal background-subtracted counts.
    All uncertainties are given at the $1\sigma$ level ($\sim$\,68\% Gaussian probability content).
    The same modulation factor $\mu$ is used for all calculations.
    }
    \label{tab:polValues}
\end{table*}

\begin{figure}
  \centering
  \includegraphics[width=0.475\textwidth]{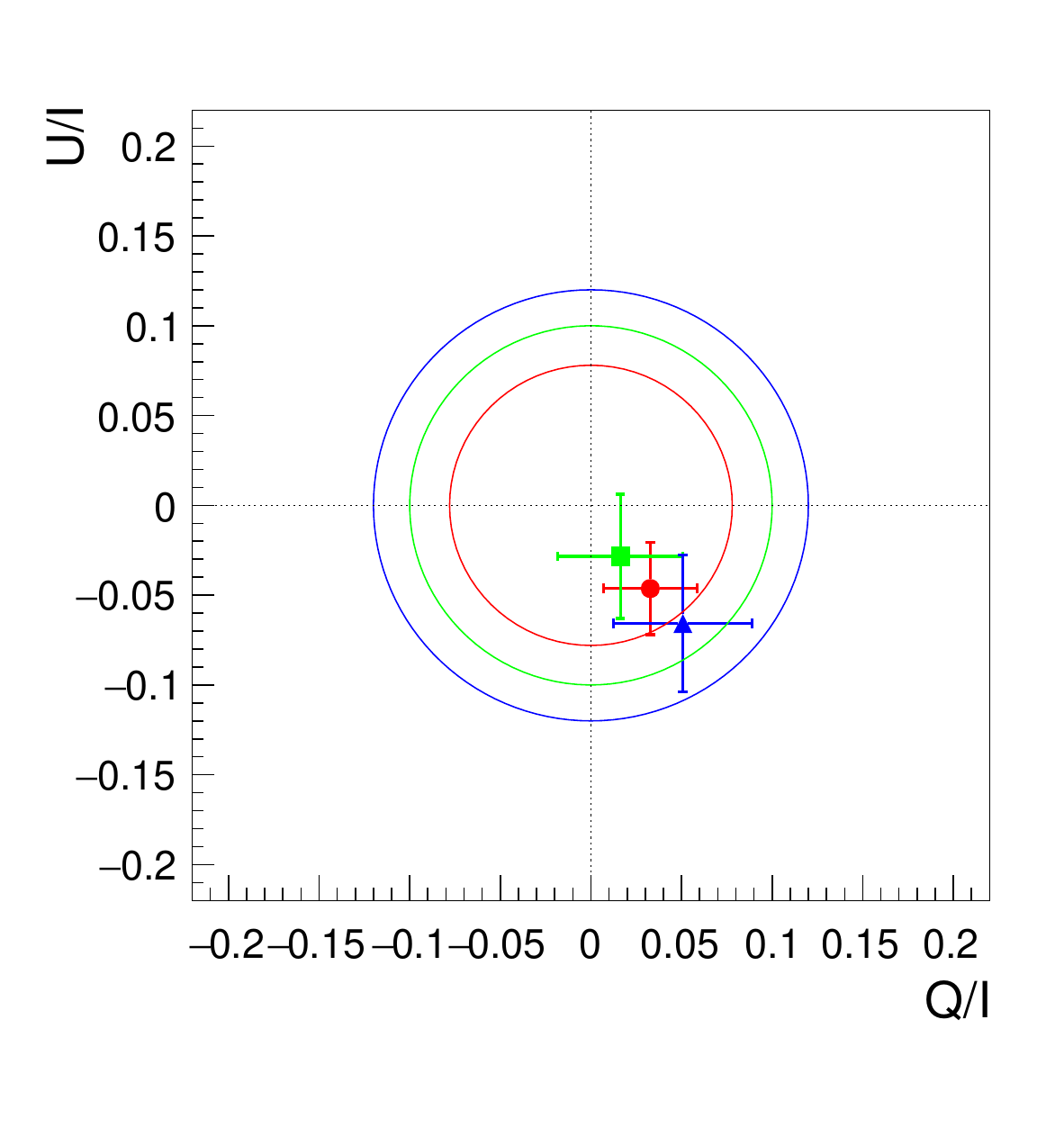}
  \caption{
  The $\sim$\,19\,--\,64\,keV (red dot), $\sim$\,19\,--\,35\,keV (green square), and $\sim$\,35\,--\,64\,keV (blue triangle) background-subtracted, normalized Stokes results, with errors bars giving the 1$\sigma$ confidence level. 
  The circles delineate the $MDP_{99}$ values (also given in Table \ref{tab:polValues}) of each of these energy ranges (inner circle $\sim$\,19\,--\,64\,keV, middle circle $\sim$\,19\,--\,35\,keV, and outer circle $\sim$\,35\,--\,64\,keV; all energy ranges as recorded in the CZT detectors).
  }
  \label{fig:stokesPlane}
\end{figure}

The normalized Stokes parameters of the signal are calculated to be
$Q/I = 0.033 \,\pm\, 0.026$ and $U/I = -0.046 \,\pm\, 0.026$.
These results for the entire $\sim$\,19\,--\,64\,keV energy range, as well as results for the $\sim$\,19\,--\,35\,keV and $\sim$\,35\,--\,64\,keV energy ranges, are shown in Fig.\,\ref{fig:stokesPlane} and listed in Table\,\ref{tab:polValues}.

\begin{figure}
  \centering
  \includegraphics[width=0.475\textwidth]{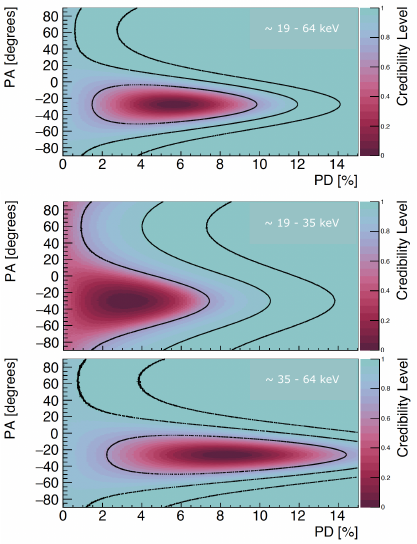}
  \caption{
  Results from the Bayesian analysis of the {\it XL-Calibur} data.
  Contour lines denote the $1\sigma$, $2\sigma$, and $3\sigma$ credibility regions. 
  The color/darkness scale denotes the lowest-percentile credibility level into which each region could fall.
  }
  \label{fig:bayesComparison}
\end{figure}

Fig.\,\ref{fig:bayesComparison} displays the results from the Bayesian analysis, showing the posterior distributions of the PDs and PAs.
We infer a marginalized $\sim$\,19\,--\,64\,keV PD of ($5.0^{+2.7}_{-3.0}$)\% 
and PA of $-28^{\circ} \pm 17^{\circ}$
(Table\,\ref{tab:polValues}), with uncertainties determined by the 1$\sigma$ intervals of the marginalized distributions.

We derive a concrete measure of statistical significance for the signal by making use of the normalized Stokes parameters, $q=Q/I$ and $u=U/I$, and their 1$\sigma$ errors, $\sigma_q$ and $\sigma_u$. 
Using the statistical distance,  $d\,=\,\sqrt{(q/\sigma_q)^2+(u/\sigma_u)^2}$, between a polarization measurement and an unpolarized beam,
the probability $p_c(d)$ for obtaining a larger 
PD by chance is given by
\begin{eqnarray*}
p_c(d)\,=\,\frac{1}{2\pi}\int_d^{\infty} 2\pi r dr\,e^{-r^2/2}.
\end{eqnarray*}
For the overall Cyg X-1 result, we thus infer a chance probability of $p_c\,=\,0.087$ for detecting larger PDs than the one we observed, given an unpolarized signal.
This measure of significance confirms, more rigorously, the approximate significance that one would nominally obtain through dividing the measured PD value in Table\,\ref{tab:polValues} by its corresponding error.


\section{Discussion and Summary}
\label{s:discussion}

\begin{figure}
  \centering
  \includegraphics[width=0.40\textwidth]{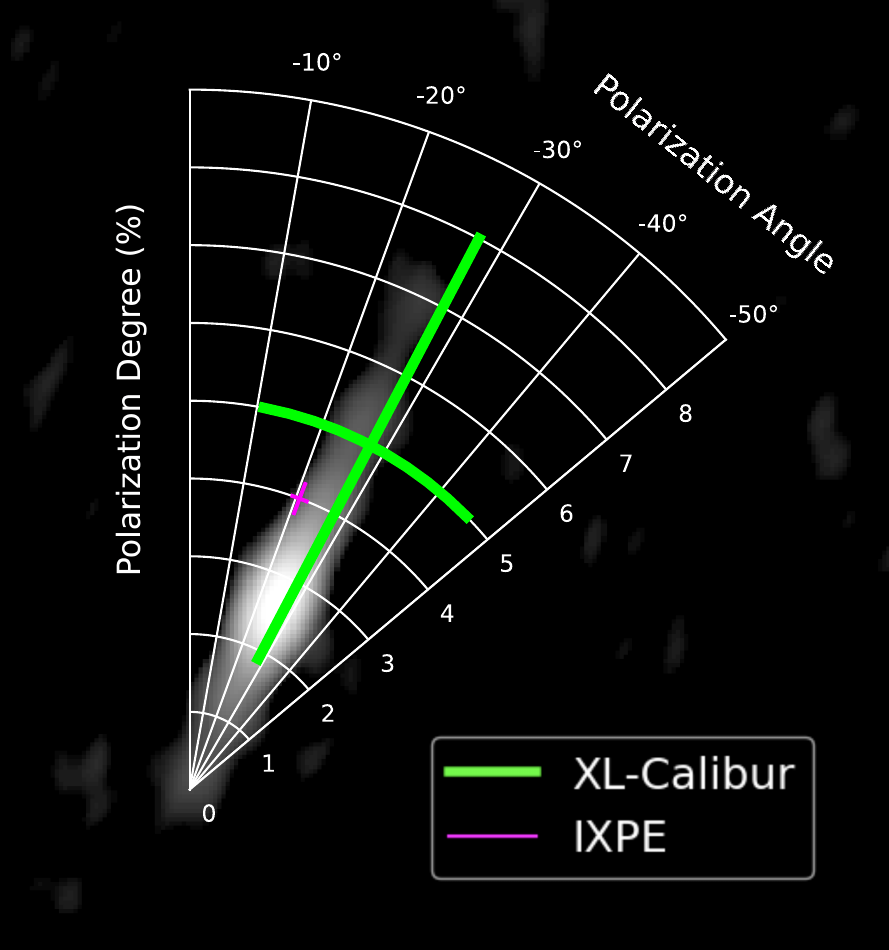}
  \caption{
  The polar diagram of measured PD and PA values in the X-ray band overlaid atop the radio jet image from VLBA astrometry data \citep{2021Sci...371.1046M}; PD values are in percent in the radial direction.
  Results are given for the {\it XL-Calibur} (green) observation across the entire $\sim$\,19\,--\,64\,keV energy range, as well as for the {\it IXPE} (magenta) results across its 2\,--\,8\,keV band.
  The errors bars of the polarization results are given at the 1$\sigma$ level, as derived from the marginalized Bayesian analysis (for {\it XL-Calibur}) and as listed in Table S2 of \citet{2022Sci...378..650K} (for {\it IXPE}). 
  }
  \label{fig:jetOverlay}
\end{figure}

{\it XL-Calibur} observed Cyg X-1 in the low/hard state in July 2024. 
The observations revealed a $\sim$\,19\,--\,64\,keV PD of ($5.0^{+2.7}_{-3.0}$)\% at a PA of $-28^{\circ} \pm 17^{\circ}$, 
similar to the 2\,--\,8\,keV PD of (4.01$\pm$0.20)\% and PA of $-20^{\circ}.7 \pm 1^{\circ}.4$ measured by {\it IXPE}. 
The results are also consistent with the previous best measurement, from {\it PoGO+} \citep{2018NatAs...2..652C}, compared to which {\it XL-Calibur} offers around a factor of two reduction in the PD error.  
The measured PA of the hard X-ray emission, just as for the soft X-ray results from {\it IXPE}, aligns with the radio jet as determined from Very Long Baseline Array (VLBA) observations at 8.4 GHz \citep{2021Sci...371.1046M}. 
The overlay of the projection of the polarization results from both bandpasses atop the radio jet image is shown in Fig.\,\ref{fig:jetOverlay}.

\begin{figure}
  \centering
  \includegraphics[width=0.475\textwidth]{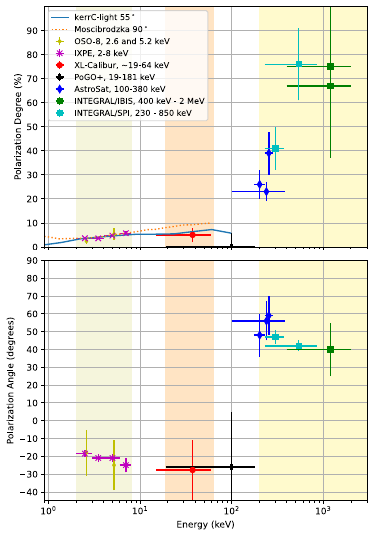}
  \caption{Compilation of Cyg X-1 polarization results from the X-ray band to the gamma-ray band. The shaded regions show the 2\,--\,8\,keV energy range of {\it IXPE} (left), the $\sim$\,19\,--\,64\,keV energy range of {\it XL-Calibur} (middle), and the 200\,keV\,--\,5\,MeV energy range of the  upcoming {\it Compton Spectrometer and Imager} ({\it COSI}) (right) \citep{2024icrc.confE.745T}.
  The multiwavelength data are taken from \citet{1980ApJ...238..710L, 2022Sci...378..650K, 2018NatAs...2..652C, 2024ApJ...960L...2C, 2011Sci...332..438L, 2012ApJ...761...27J, 2015ApJ...807...17R}.
  All of these measurements observe Cyg X-1 in some hard (or intermediate-hard) state \citep{2024ApJ...960L...2C}, except for {\it OSO-8}.
  The solid blue line delineates the PD prediction of the 55$^{\circ}$ inclination {\it kerrC}-light model (see text; see also \citet{2022ApJ...934....4K}).
  The dotted orange line delineates the PD prediction of the 90$^{\circ}$ model from \citet{2024Ap&SS.369...68M}.
  Both of these models predict a slight increase of the PD and a roughly constant PA, parallel to the jet, across the {\it IXPE} and {\it XL-Calibur} energy ranges.
  }
  \label{fig:comparisonPDPA}
\end{figure}

\begin{figure}
  \centering
  \includegraphics[width=0.475\textwidth]{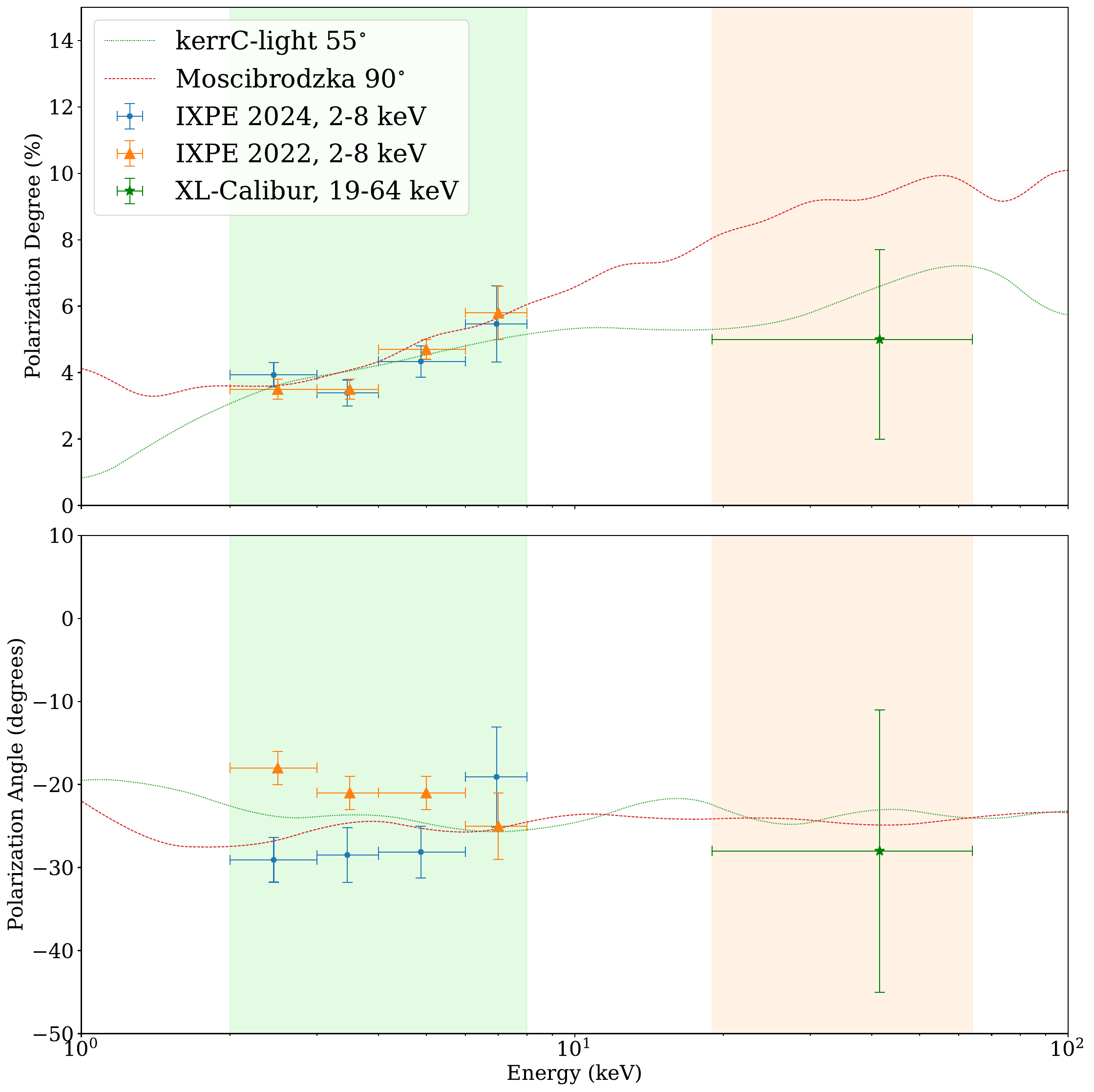}
  \caption{
  A zoom-in on the models and data in the X-ray band. 
  The {\it kerrC} model is a Monte Carlo-based model: 68,040 combinations of system parameters were simulated, with 20,000,000 photons used for each configuration.
  The {\it IXPE} 2024 data come from {\it IXPE} observations 03010001, 03010101, and 03003101 of Cyg X-1, which were taken within a couple months before the {\it XL-Calibur} observation, and are analyzed jointly with {\it ixpeobssim} \citep{2022SoftX..1901194B}. 
  }
  \label{fig:pcube}
\end{figure}

Fig.\,\ref{fig:comparisonPDPA} shows the {\it XL-Calibur} data in the context of polarization results obtained across the X-ray and $\gamma$-ray bands.
The low levels of polarization and comparability of the measured PDs and PAs from 2 keV to $\sim$\,64 keV 
suggest that similar physical mechanisms are responsible for the polarization properties observed in both bands.
The $\sim$\,19\,--\,64\,keV X-ray emission from BHXRBs is thought to be dominated by Comptonized emission from the hot coronal plasma, some of which reaches the observer after reflecting at least once off the accretion disk.
However, above 100 keV, {\it AstroSat} and {\it INTEGRAL} measurements indicate a drastic rise of the PD with respect to energy.
Indeed, spectral analyses show that the high-energy emission transitions from being corona-dominated (irrespective of the specific coronal emission model being used to explain the hard X-rays) to being jet-dominated between 100\,keV and 1\,MeV \citep[see][and references therein]{2021MNRAS.500.2112K}.
Moreover, the PA seems to swing by $\sim$\,90$^{\circ}$ between the {\it XL-Calibur} and {\it AstroSat} energy ranges, as would be expected for a transition from emission Comptonized in a horizontally-extended corona to synchrotron and/or inverse Compton emission from a jet with an axial magnetic field. 
However, the {\it AstroSat} and {\it INTEGRAL} results were obtained from instruments not specifically built for polarimetry, leading to larger systematic errors on the results. The {\it COSI} mission, a large and uniform Compton telescope for polarimetry to be launched in 2027, should help resolve this question with precise polarization measurements in its energy range of 200\,keV\,--\,5\,MeV \citep{2022hxga.book...73T}.

\begin{figure}
  \centering
  \includegraphics[width=0.475\textwidth]{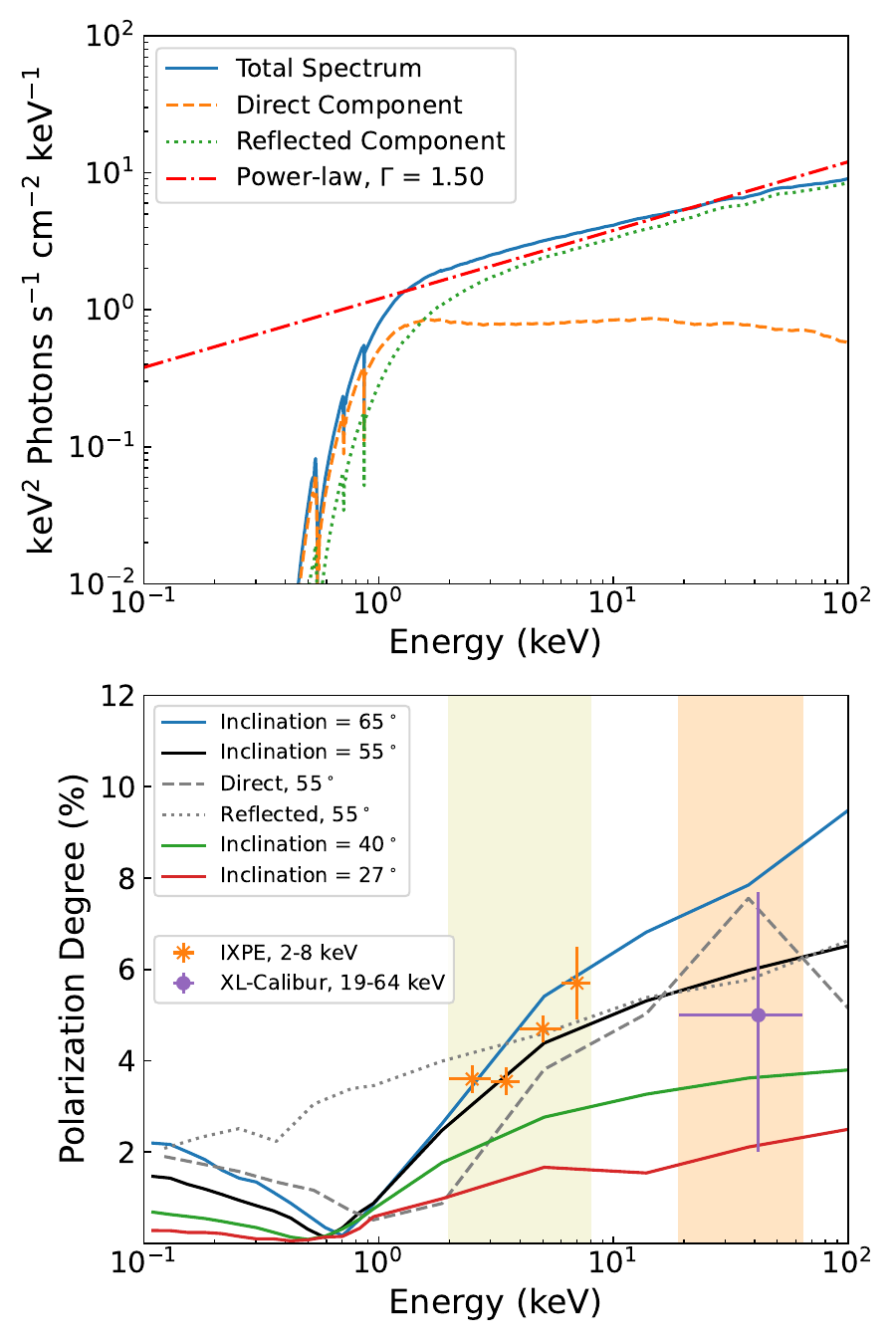}
  \caption{
  Top: 
  Simulated Cyg X-1 energy spectrum from the {\it kerrC-light} code, assuming a geometrically thin, optically thick accretion disk with a sandwich corona of opening angle 10$^{\circ}$, of optical depth $\tau\,=\,$0.41, and of inclination angle 55$^\circ$.
  While the dashed orange line shows the flux of the photons that reach the observer without any scattering off the disk, the dotted green line shows the flux of the photons scattering off the disk one or several times. 
  A dotted-dashed red line is also presented to indicate the power-law index calculated in Fig.\,\ref{fig:cygnusCrabRatio}.
  Bottom: 
  Polarization degree spectra of the {\it kerrC}-light model from the above panel, varying only the inclination in each instance (with the model from the highest inclination having the highest PDs). 
  The {\it IXPE} (from 2022) and {\it XL-Calibur} data are overlaid, as in Figs.\,\ref{fig:comparisonPDPA} and \ref{fig:pcube}. 
  PD as a function of energy, for the direct and reflected components, is also plotted for the 55$^\circ$ case as dashed and dotted grey lines.
  The abrupt changes in the flux of the direct component at higher energies reflects the low statistics due to the low proportion of flux it comprises at those energies. 
  \label{fig:kerrc}}
\end{figure}

Focusing in on the X-ray band: Figs.\,\ref{fig:comparisonPDPA} and \ref{fig:pcube} compare the {\it IXPE}, {\it PoGO+}, and {\it XL-Calibur} results with various published theoretical estimates. 
The prediction of the {\it kerrC}-light model \citep{2022ApJ...934....4K}, given by the solid line in Fig.\,\ref{fig:comparisonPDPA}, assumes a wedge-shaped corona sandwiching a thin, 100\%-reflecting accretion disk.  
The reflection is modeled using Chandrasekhar's classical results for an infinitely deep electron scattering atmosphere \citep{1960ratr.book.....C}.
This model has the same parameter values as shown in Table S3 of \cite{2022Sci...378..650K}, save for a dimensionless spin parameter $a$ ($-$1\,$\le a\le $\,1) of 0.94 (as measured by \citet{2016ApJ...826...87W}), an increased accretion rate (for flux normalization purposes), and an inclination angle of 55$^{\circ}$ (the angle that best fits both soft and hard X-ray data, as shown in Fig.\,\ref{fig:kerrc}, given these parameters).
Such a model predicts a slight increase of PD (from $\sim$\,2\% to $\sim$\,6\%) when going from soft to hard X-rays, as well as a stable PA, parallel to the black hole spin axis.
This model is thus one model that is consistent with both the {\it IXPE} and {\it XL-Calibur} results.

The top panel of Fig.\,\ref{fig:kerrc} offers a closer look at this model of inclination 55$^\circ$, showing the total predicted energy spectrum, as well as the spectrum subdivided into `direct' and `reflected' contributions. 
The accretion disk emits photons of about a few keV in energy, and these photons gain more energy the more often they scatter in the corona. 
For a sandwich corona, a larger number of scatterings for a photon (that is, a photon more likely to be at a higher energy) is accompanied by a higher likelihood of one or more encounters and reflections off the disk, given the geometry. 
In the sandwich model, reflection off the disk thus becomes an integral part of the Comptonization process, and the fraction of reflected photons increases with increasing photon energy.  
The distinction between direct coronal emission and reflected coronal emission is thus not as meaningful as it would be for models using a more localized corona situated further away from the disk. 
The power-law index of 1.55 calculated in Fig.\,\ref{fig:cygnusCrabRatio} is overplotted in Fig.\,\ref{fig:comparisonPDPA}, displaying a reasonable agreement with the spectral flux predicted by {\it kerrC}-light in the X-ray regime. 

As a further examination of the inclination angles, the bottom panel of Fig.\,\ref{fig:kerrc} shows the PDs predicted by {\it kerrC}-light models at several possible inclinations of the black hole. 
The PDs of the direct and reflected components of the $i=55^\circ$ model are plotted, as mentioned above, as well. 
As expected, the PD of the overall model starts very closely following the PD of the reflected component at about the same energy that the reflected flux becomes dominant in the top panel. 
Models with inclinations $<40^\circ$ have difficulty explaining both the observed {\it IXPE} and {\it XL-Calibur} PDs. 
For example, for an inclination of $i=27^\circ$, which is close to the inferred inclination angle of the binary \citep{2021Sci...371.1046M}
the sandwich model of {\it kerrC}-light maintains a PD below 3\% through the end of the {\it XL-Calibur} range.  
However, \citet{Beloborodov_1999, 2023ApJ...949L..10P} emphasize that higher PDs can result from coronal plasma outflowing at $\sim$\,40\% the speed of light or faster, thus allowing for lower inclination angles.

As mentioned in the introduction, several alternative, but related, models have also been proposed in the literature. 
In the model of \citet{2024Ap&SS.369...68M}, the X-rays come from the inverse Compton scattering of Bremsstrahlung and synchrotron emission produced in the outflowing jet wall.
The high-inclination ($90^{\circ}$) result from their General Relativistic {\it Radiative} Magnetohydrodynamic simulations reproduces the observed alignment of the soft and hard X-ray PAs with the radio jet, as well as generally tracking the change in PD between the lower and higher energy bands (as displayed in Figs.\,\ref{fig:comparisonPDPA} and \ref{fig:pcube}).
Similarly, \citet{2024MNRAS.528L.157D} argue that the Cyg X-1 PDs and PAs can be explained by the inverse Compton scattering of disk emission by a mildly-relativistic ($\sim$\,75\% of the speed of light) cold outflow in the shape of a hollow cone, centered on the black hole spin axis, into which the observer looks. 
\cite{2025ApJ...979..199S}, using General Relativistic {\it Resistive} Magnetohydrodynamic simulations, give a physical scenario which could lead to such a configuration: 
magnetic reconnection 
producing cold plasmoids, with bulk motion, that inverse Compton scatter longer-wavelength photons into the {\it IXPE} and {\it XL-Calibur} energy bands.
All of these models predict little variation of the PD and PA, provided that the {\it IXPE} signal is not heavily contaminated by disk emission.

Although the present data starts to provide some indication, a final decision between models of hard X-ray emission from BHXRBs will require both 
(i) additional observations of BHXRBs in the hard and soft states with the best possible broadband (measuring both soft and hard X-rays) spectropolarimetric coverage, as well as 
(ii) continued refinement of the various theoretical models.
Future flights of {\it XL-Calibur} will contribute to this goal through constraining the polarization from more BHXRBs in the hard X-rays.
A longer flight from McMurdo in Antarctica, observing southern-hemisphere sources, such as GX 339-4, 4U 1630-47, Swift J1727.8-1613, LMC X-1, and LMC X-3, will significantly increase the number of BHXRBs studied in the this energy range, as well as the precision of their polarization measurements. 

 
\section*{Acknowledgments}
{\it XL-Calibur} is a joint mission supported by NASA, JAXA, and the Swedish National Space Agency {\it Rymdstyrelsen}. 
We sincerely thank James Miller-Jones for the data used to produce the VLBA radio jet image.
We acknowledge NASA support under grant 80NSSC24K0205.
KTH authors are supported by the Swedish National Space Agency (2022-00178 and 2024-00248). MP also acknowledges funding from the Swedish Research Council {\it Vetenskapsrådet} (2021-05128).
The Washington University in St. Louis group acknowledges additional NASA support through the grants 80NSSC20K0329, 80NSSC21K1817, 80NSSC22K1291, 80NSSC22K1883, 80NSSC23K1041, and 80NSSC24K1178, as well as funding from the McDonnell Center for the Space Sciences at Washington University in St. Louis. 
The University of New Hampshire group acknowledges additional NASA support through the grants 80NSSC24K0636 and 80NSSC24K1762.
The Japanese Society for the Promotion of Science (JSPS) has supported this work through KAKENHI Grant Numbers 19H01908, 19H05609, 20H00175 (HM), 20H00178 (HM), 21K13946 (YU), 22H01277 (YM), 23H00117, and 23H00128 (HM).

\section*{Data Availability}
The XL-Calibur data underlying this article will be made available via the NASA HEASARC data archive, at https://heasarc.gsfc.nasa.gov/docs/xlcalibur/.


\bibliographystyle{aasjournal}
\bibliography{references_v2} 

\begin{thebibliography}{}
\expandafter\ifx\csname natexlab\endcsname\relax\def\natexlab#1{#1}\fi
\providecommand{\url}[1]{\href{#1}{#1}}
\providecommand{\dodoi}[1]{doi:~\href{http://doi.org/#1}{\nolinkurl{#1}}}
\providecommand{\doeprint}[1]{\href{http://ascl.net/#1}{\nolinkurl{http://ascl.net/#1}}}
\providecommand{\doarXiv}[1]{\href{https://arxiv.org/abs/#1}{\nolinkurl{https://arxiv.org/abs/#1}}}

\bibitem[{{Abarr} {et~al.}(2020){Abarr}, {Baring}, {Beheshtipour}, {Beilicke}, {de Geronimo}, {Dowkontt}, {Errando}, {Guarino}, {Iyer}, {Kislat}, {Kiss}, {Kitaguchi}, {Krawczynski}, {Lanzi}, {Li}, {Lisalda}, {Okajima}, {Pearce}, {Press}, {Rauch}, {Stuchlik}, {Takahashi}, {Tang}, {Uchida}, {West}, {Jenke}, {Krimm}, {Lien}, {Malacaria}, {Miller}, \& {Wilson-Hodge}}]{2020ApJ...891...70A}
{Abarr}, Q., {Baring}, M., {Beheshtipour}, B., {et~al.} 2020, \apj, 891, 70, \dodoi{10.3847/1538-4357/ab672c}

\bibitem[{{Abarr} {et~al.}(2021){Abarr}, {Awaki}, {Baring}, {Bose}, {De Geronimo}, {Dowkontt}, {Errando}, {Guarino}, {Hattori}, {Hayashida}, {Imazato}, {Ishida}, {Iyer}, {Kislat}, {Kiss}, {Kitaguchi}, {Krawczynski}, {Lisalda}, {Matake}, {Maeda}, {Matsumoto}, {Mineta}, {Miyazawa}, {Mizuno}, {Okajima}, {Pearce}, {Rauch}, {Ryde}, {Shreves}, {Spooner}, {Stana}, {Takahashi}, {Takeo}, {Tamagawa}, {Tamura}, {Tsunemi}, {Uchida}, {Uchida}, {West}, {Wulf}, \& {Yamamoto}}]{2021APh...12602529A}
{Abarr}, Q., {Awaki}, H., {Baring}, M.~G., {et~al.} 2021, Astroparticle Physics, 126, 102529, \dodoi{10.1016/j.astropartphys.2020.102529}

\bibitem[{{Aoyagi} {et~al.}(2024){Aoyagi}, {Bose}, {Chun}, {Gau}, {Hu}, {Ishiwata}, {Iyer}, {Kislat}, {Kiss}, {Klepper}, {Krawczynski}, {Lisalda}, {Maeda}, {Malmborg}, {Matsumoto}, {Miyamoto}, {Miyazawa}, {Pearce}, {Rauch}, {Rodriguez Cavero}, {Spooner}, {Takahashi}, {Uchida}, {West}, {Wimalasena}, \& {Yoshimoto}}]{2024APh...15802944A}
{Aoyagi}, M., {Bose}, R.~G., {Chun}, S., {et~al.} 2024, Astroparticle Physics, 158, 102944, \dodoi{10.1016/j.astropartphys.2024.102944}

\bibitem[{{Awaki} {et~al.}(2014){Awaki}, {Kunieda}, {Ishida}, {Matsumoto}, {Babazaki}, {Demoto}, {Furuzawa}, {Haba}, {Hayashi}, {Iizuka}, {Ishibashi}, {Ishida}, {Itoh}, {Iwase}, {Kosaka}, {Kurihara}, {Kuroda}, {Maeda}, {Meshino}, {Mitsuishi}, {Miyata}, {Miyazawa}, {Mori}, {Nagano}, {Namba}, {Ogasaka}, {Ogi}, {Okajima}, {Saji}, {Shimasaki}, {Sato}, {Sato}, {Sugita}, {Suzuki}, {Tachibana}, {Tachibana}, {Takizawa}, {Tamura}, {Tawara}, {Torii}, {Uesugi}, {Yamashita}, \& {Yamauchi}}]{2014ApOpt..53.7664A}
{Awaki}, H., {Kunieda}, H., {Ishida}, M., {et~al.} 2014, \ao, 53, 7664, \dodoi{10.1364/AO.53.007664}

\bibitem[{{Awaki} {et~al.}(2025){Awaki}, {Baring}, {Bose}, {Braun}, {Casey}, {Chun}, {Galchenko}, {Gau}, {Goya}, {Hakamata}, {Hayashi}, {Heatwole}, {Hu}, {Imazawa}, {Ishi}, {Ishida}, {Kislat}, {Kiss}, {Klepper}, {Krawczynski}, {Kuramoto}, {Lanzi}, {Lisalda}, {Maeda}, {af Malmborg}, {Matsumoto}, {Menon}, {Miyamoto}, {Miyamoto}, {Miyazawa}, {Murakami}, {Nagao}, {Okajima}, {Pearce}, {Rauch}, {Cavero}, {Shima}, {Shirahama}, {Snow}, {Spooner}, {Takahashi}, {Takatsuka}, {Tamura}, {Tanaka}, {Uchida}, {West}, {Wulf}, {Yokota}, \& {Yoshimoto}}]{2025MNRAS.tmpL..25A}
{Awaki}, H., {Baring}, M.~G., {Bose}, R., {et~al.} 2025, \mnras, \dodoi{10.1093/mnrasl/slaf026}

\bibitem[{{Baldini} {et~al.}(2022){Baldini}, {Bucciantini}, {Lalla}, {Ehlert}, {Manfreda}, {Negro}, {Omodei}, {Pesce-Rollins}, {Sgr{\`o}}, \& {Silvestri}}]{2022SoftX..1901194B}
{Baldini}, L., {Bucciantini}, N., {Lalla}, N.~D., {et~al.} 2022, SoftwareX, 19, 101194, \dodoi{10.1016/j.softx.2022.101194}

\bibitem[{{Begelman} \& {Sikora}(1987)}]{1987ApJ...322..650B}
{Begelman}, M.~C., \& {Sikora}, M. 1987, \apj, 322, 650, \dodoi{10.1086/165760}

\bibitem[{Beloborodov(1998)}]{Beloborodov_1999}
Beloborodov, A.~M. 1998, The Astrophysical Journal, 510, L123, \dodoi{10.1086/311810}

\bibitem[{{Beloborodov}(2017)}]{2017ApJ...850..141B}
{Beloborodov}, A.~M. 2017, \apj, 850, 141, \dodoi{10.3847/1538-4357/aa8f4f}

\bibitem[{{Cavero} \& {IXPE Collaboration}(2024)}]{2024mbhe.confE..39C}
{Cavero}, N.~R., \& {IXPE Collaboration}. 2024, in Multifrequency Behaviour of High Energy Cosmic Sources XIV, 39

\bibitem[{{Chandrasekhar}(1960)}]{1960ratr.book.....C}
{Chandrasekhar}, S. 1960, {Radiative transfer}

\bibitem[{{Chattopadhyay} {et~al.}(2024){Chattopadhyay}, {Kumar}, {Rao}, {Bhargava}, {Vadawale}, {Ratheesh}, {Dewangan}, {Bhattacharya}, {Mithun}, \& {Bhalerao}}]{2024ApJ...960L...2C}
{Chattopadhyay}, T., {Kumar}, A., {Rao}, A.~R., {et~al.} 2024, \apjl, 960, L2, \dodoi{10.3847/2041-8213/ad118d}

\bibitem[{{Chauvin} {et~al.}(2017){Chauvin}, {Flor{\'e}n}, {Friis}, {Jackson}, {Kamae}, {Kataoka}, {Kawano}, {Kiss}, {Mikhalev}, {Mizuno}, {Ohashi}, {Stana}, {Tajima}, {Takahashi}, {Uchida}, \& {Pearce}}]{2017NatSR...7.7816C}
{Chauvin}, M., {Flor{\'e}n}, H.~G., {Friis}, M., {et~al.} 2017, Scientific Reports, 7, 7816, \dodoi{10.1038/s41598-017-07390-7}

\bibitem[{{Chauvin} {et~al.}(2018){Chauvin}, {Flor{\'e}n}, {Friis}, {Jackson}, {Kamae}, {Kataoka}, {Kawano}, {Kiss}, {Mikhalev}, {Mizuno}, {Ohashi}, {Stana}, {Tajima}, {Takahashi}, {Uchida}, \& {Pearce}}]{2018NatAs...2..652C}
---. 2018, Nature Astronomy, 2, 652, \dodoi{10.1038/s41550-018-0489-x}

\bibitem[{{Dexter} \& {Begelman}(2024)}]{2024MNRAS.528L.157D}
{Dexter}, J., \& {Begelman}, M.~C. 2024, \mnras, 528, L157, \dodoi{10.1093/mnrasl/slad182}

\bibitem[{{Doroshenko} {et~al.}(2022){Doroshenko}, {Poutanen}, {Tsygankov}, {Suleimanov}, {Bachetti}, {Caiazzo}, {Costa}, {Di Marco}, {Heyl}, {La Monaca}, {Muleri}, {Mushtukov}, {Pavlov}, {Ramsey}, {Rankin}, {Santangelo}, {Soffitta}, {Staubert}, {Weisskopf}, {Zane}, {Agudo}, {Antonelli}, {Baldini}, {Baumgartner}, {Bellazzini}, {Bianchi}, {Bongiorno}, {Bonino}, {Brez}, {Bucciantini}, {Capitanio}, {Castellano}, {Cavazzuti}, {Ciprini}, {De Rosa}, {Del Monte}, {Di Gesu}, {Di Lalla}, {Donnarumma}, {Dov{\v{c}}iak}, {Ehlert}, {Enoto}, {Evangelista}, {Fabiani}, {Ferrazzoli}, {Garcia}, {Gunji}, {Hayashida}, {Iwakiri}, {Jorstad}, {Karas}, {Kitaguchi}, {Kolodziejczak}, {Krawczynski}, {Latronico}, {Liodakis}, {Maldera}, {Manfreda}, {Marin}, {Marinucci}, {Marscher}, {Marshall}, {Matt}, {Mitsuishi}, {Mizuno}, {Ng}, {O'Dell}, {Omodei}, {Oppedisano}, {Papitto}, {Peirson}, {Perri}, {Pesce-Rollins}, {Pilia}, {Possenti}, {Puccetti}, {Ratheesh}, {Romani}, {Sgr{\`o}}, {Slane}, {Spandre}, {Sunyaev}, {Tamagawa}, {Tavecchio},
  {Taverna}, {Tawara}, {Tennant}, {Thomas}, {Tombesi}, {Trois}, {Turolla}, {Vink}, {Wu}, \& {Xie}}]{2022NatAs...6.1433D}
{Doroshenko}, V., {Poutanen}, J., {Tsygankov}, S.~S., {et~al.} 2022, Nature Astronomy, 6, 1433, \dodoi{10.1038/s41550-022-01799-5}

\bibitem[{{Dov{\v{c}}iak} {et~al.}(2024){Dov{\v{c}}iak}, {Podgorn{\'y}}, {Svoboda}, {Steiner}, {Kaaret}, {Krawczynski}, {Ingram}, {Kravtsov}, {Marra}, {Muleri}, {Garc{\'\i}a}, {Mastroserio}, {Miku{\v{s}}incov{\'a}}, {Ratheesh}, \& {Cavero}}]{2024Galax..12...54D}
{Dov{\v{c}}iak}, M., {Podgorn{\'y}}, J., {Svoboda}, J., {et~al.} 2024, Galaxies, 12, 54, \dodoi{10.3390/galaxies12050054}

\bibitem[{{Ehlert} {et~al.}(2022){Ehlert}, {Ferrazzoli}, {Marinucci}, {Marshall}, {Middei}, {Pacciani}, {Perri}, {Petrucci}, {Puccetti}, {Barnouin}, {Bianchi}, {Liodakis}, {Madejski}, {Marin}, {Marscher}, {Matt}, {Poutanen}, {Wu}, {Agudo}, {Antonelli}, {Bachetti}, {Baldini}, {Baumgartner}, {Bellazzini}, {Bongiorno}, {Bonino}, {Brez}, {Bucciantini}, {Capitanio}, {Castellano}, {Cavazzuti}, {Ciprini}, {Costa}, {De Rosa}, {Del Monte}, {Di Gesu}, {Di Lalla}, {Di Marco}, {Donnarumma}, {Doroshenko}, {Dov{\v{c}}iak}, {Enoto}, {Evangelista}, {Fabiani}, {Garcia}, {Gunji}, {Hayashida}, {Heyl}, {Iwakiri}, {Jorstad}, {Karas}, {Kitaguchi}, {Kolodziejczak}, {Krawczynski}, {La Monaca}, {Latronico}, {Maldera}, {Manfreda}, {Massaro}, {Mitsuishi}, {Mizuno}, {Muleri}, {Negro}, {Ng}, {O'Dell}, {Omodei}, {Oppedisano}, {Papitto}, {Pavlov}, {Peirson}, {Pesce-Rollins}, {Pilia}, {Possenti}, {Ramsey}, {Rankin}, {Ratheesh}, {Romani}, {Sgr{\`o}}, {Slane}, {Soffitta}, {Spandre}, {Tamagawa}, {Tavecchio}, {Taverna}, {Tawara}, {Tennant},
  {Thomas}, {Tombesi}, {Trois}, {Tsygankov}, {Turolla}, {Vink}, {Weisskopf}, {Xie}, {Zane}, {IXPE Collaboration}, {Rodi}, {Jourdain}, \& {Roques}}]{2022ApJ...935..116E}
{Ehlert}, S.~R., {Ferrazzoli}, R., {Marinucci}, A., {et~al.} 2022, \apj, 935, 116, \dodoi{10.3847/1538-4357/ac8056}

\bibitem[{{Galchenko} \& {Pernicka}(2022)}]{2022JGCD...45.1365G}
{Galchenko}, P., \& {Pernicka}, H. 2022, Journal of Guidance Control Dynamics, 45, 1365, \dodoi{10.2514/1.G006465}

\bibitem[{{Gianolli} {et~al.}(2023){Gianolli}, {Kim}, {Bianchi}, {Ag{\'\i}s-Gonz{\'a}lez}, {Madejski}, {Marin}, {Marinucci}, {Matt}, {Middei}, {Petrucci}, {Soffitta}, {Tagliacozzo}, {Tombesi}, {Ursini}, {Barnouin}, {De Rosa}, {Di Gesu}, {Ingram}, {Loktev}, {Panagiotou}, {Podgorny}, {Poutanen}, {Puccetti}, {Ratheesh}, {Veledina}, {Zhang}, {Agudo}, {Antonelli}, {Bachetti}, {Baldini}, {Baumgartner}, {Bellazzini}, {Bongiorno}, {Bonino}, {Brez}, {Bucciantini}, {Capitanio}, {Castellano}, {Cavazzuti}, {Chen}, {Ciprini}, {Costa}, {Del Monte}, {Di Lalla}, {Di Marco}, {Donnarumma}, {Doroshenko}, {Dov{\v{c}}iak}, {Ehlert}, {Enoto}, {Evangelista}, {Fabiani}, {Ferrazzoli}, {Garc{\'\i}a}, {Gunji}, {Heyl}, {Iwakiri}, {Jorstad}, {Kaaret}, {Karas}, {Kislat}, {Kitaguchi}, {Kolodziejczak}, {Krawczynski}, {La Monaca}, {Latronico}, {Liodakis}, {Maldera}, {Manfreda}, {Marscher}, {Marshall}, {Massaro}, {Mitsuishi}, {Mizuno}, {Muleri}, {Negro}, {Ng}, {O'Dell}, {Omodei}, {Oppedisano}, {Papitto}, {Pavlov}, {Peirson}, {Perri},
  {Pesce-Rollins}, {Pilia}, {Possenti}, {Ramsey}, {Rankin}, {Roberts}, {Romani}, {Sgr{\`o}}, {Slane}, {Spandre}, {Swartz}, {Tamagawa}, {Tavecchio}, {Taverna}, {Tawara}, {Tennant}, {Thomas}, {Trois}, {Tsygankov}, {Turolla}, {Vink}, {Weisskopf}, {Wu}, {Xie}, \& {Zane}}]{2023MNRAS.523.4468G}
{Gianolli}, V.~E., {Kim}, D.~E., {Bianchi}, S., {et~al.} 2023, \mnras, 523, 4468, \dodoi{10.1093/mnras/stad1697}

\bibitem[{{Gianolli} {et~al.}(2024){Gianolli}, {Bianchi}, {Kammoun}, {Gnarini}, {Marinucci}, {Ursini}, {Parra}, {Tortosa}, {De Rosa}, {Kim}, {Marin}, {Matt}, {Serafinelli}, {Soffitta}, {Tagliacozzo}, {Di Gesu}, {Done}, {Marshall}, {Middei}, {Mikusincova}, {Petrucci}, {Ravi}, {Svoboda}, \& {Tombesi}}]{2024A&A...691A..29G}
{Gianolli}, V.~E., {Bianchi}, S., {Kammoun}, E., {et~al.} 2024, \aap, 691, A29, \dodoi{10.1051/0004-6361/202451645}

\bibitem[{{Gierli{\'n}ski} \& {Zdziarski}(1999)}]{1999ASPC..161...64G}
{Gierli{\'n}ski}, M., \& {Zdziarski}, A.~A. 1999, in Astronomical Society of the Pacific Conference Series, Vol. 161, High Energy Processes in Accreting Black Holes, ed. J.~{Poutanen} \& R.~{Svensson}, 64, \dodoi{10.48550/arXiv.astro-ph/9811220}

\bibitem[{{Grinberg} {et~al.}(2013){Grinberg}, {Hell}, {Pottschmidt}, {B{\"o}ck}, {Nowak}, {Rodriguez}, {Bodaghee}, {Cadolle Bel}, {Case}, {Hanke}, {K{\"u}hnel}, {Markoff}, {Pooley}, {Rothschild}, {Tomsick}, {Wilson-Hodge}, \& {Wilms}}]{2013A&A...554A..88G}
{Grinberg}, V., {Hell}, N., {Pottschmidt}, K., {et~al.} 2013, \aap, 554, A88, \dodoi{10.1051/0004-6361/201321128}

\bibitem[{{Ingram} {et~al.}(2023){Ingram}, {Ewing}, {Marinucci}, {Tagliacozzo}, {Rosario}, {Veledina}, {Kim}, {Marin}, {Bianchi}, {Poutanen}, {Matt}, {Marshall}, {Ursini}, {De Rosa}, {Petrucci}, {Madejski}, {Barnouin}, {Gesu}, {Dov{\v{c}}iak}, {Gianolli}, {Krawczynski}, {Loktev}, {Middei}, {Podgorny}, {Puccetti}, {Ratheesh}, {Soffitta}, {Tombesi}, {Ehlert}, {Massaro}, {Agudo}, {Antonelli}, {Bachetti}, {Baldini}, {Baumgartner}, {Bellazzini}, {Bongiorno}, {Bonino}, {Brez}, {Bucciantini}, {Capitanio}, {Castellano}, {Cavazzuti}, {Chen}, {Ciprini}, {Costa}, {Del Monte}, {Lalla}, {Marco}, {Donnarumma}, {Doroshenko}, {Enoto}, {Evangelista}, {Fabiani}, {Ferrazzoli}, {Garc{\'\i}a}, {Gunji}, {Heyl}, {Iwakiri}, {Jorstad}, {Kaaret}, {Karas}, {Kislat}, {Kitaguchi}, {Kolodziejczak}, {Monaca}, {Latronico}, {Liodakis}, {Maldera}, {Manfreda}, {Marscher}, {Mitsuishi}, {Mizuno}, {Muleri}, {Negro}, {Ng}, {O'Dell}, {Omodei}, {Oppedisano}, {Papitto}, {Pavlov}, {Peirson}, {Perri}, {Pesce-Rollins}, {Pilia}, {Possenti}, {Ramsey},
  {Rankin}, {Roberts}, {Romani}, {Sgr{\`o}}, {Slane}, {Spandre}, {Swartz}, {Tamagawa}, {Tavecchio}, {Taverna}, {Tawara}, {Tennant}, {Thomas}, {Trois}, {Tsygankov}, {Turolla}, {Vink}, {Weisskopf}, {Wu}, {Xie}, \& {Zane}}]{2023MNRAS.525.5437I}
{Ingram}, A., {Ewing}, M., {Marinucci}, A., {et~al.} 2023, \mnras, 525, 5437, \dodoi{10.1093/mnras/stad2625}

\bibitem[{{Ingram} {et~al.}(2024){Ingram}, {Bollemeijer}, {Veledina}, {Dov{\v{c}}iak}, {Poutanen}, {Egron}, {Russell}, {Trushkin}, {Negro}, {Ratheesh}, {Capitanio}, {Connors}, {Neilsen}, {Kraus}, {Iacolina}, {Pellizzoni}, {Pilia}, {Carotenuto}, {Matt}, {Mastroserio}, {Kaaret}, {Bianchi}, {Garc{\'\i}a}, {Bachetti}, {Wu}, {Costa}, {Ewing}, {Kravtsov}, {Krawczynski}, {Loktev}, {Marinucci}, {Marra}, {Miku{\v{s}}incov{\'a}}, {Nathan}, {Parra}, {Petrucci}, {Righini}, {Soffitta}, {Steiner}, {Svoboda}, {Tombesi}, {Tugliani}, {Ursini}, {Yang}, {Zane}, {Zhang}, {Agudo}, {Antonelli}, {Baldini}, {Baumgartner}, {Bellazzini}, {Bongiorno}, {Bonino}, {Brez}, {Bucciantini}, {Castellano}, {Cavazzuti}, {Chen}, {Ciprini}, {De Rosa}, {Del Monte}, {Di Gesu}, {Di Lalla}, {Di Marco}, {Donnarumma}, {Doroshenko}, {Ehlert}, {Enoto}, {Evangelista}, {Fabiani}, {Ferrazzoli}, {Gunji}, {Hayashida}, {Heyl}, {Iwakiri}, {Jorstad}, {Karas}, {Kislat}, {Kitaguchi}, {Kolodziejczak}, {La Monaca}, {Latronico}, {Liodakis}, {Maldera}, {Manfreda},
  {Marin}, {Marscher}, {Marshall}, {Massaro}, {Mitsuishi}, {Mizuno}, {Muleri}, {Ng}, {O'Dell}, {Omodei}, {Oppedisano}, {Papitto}, {Pavlov}, {Peirson}, {Perri}, {Pesce-Rollins}, {Possenti}, {Puccetti}, {Ramsey}, {Rankin}, {Roberts}, {Romani}, {Sgr{\`o}}, {Slane}, {Spandre}, {Swartz}, {Tamagawa}, {Tavecchio}, {Taverna}, {Tawara}, {Tennant}, {Thomas}, {Trois}, {Tsygankov}, {Turolla}, {Vink}, {Weisskopf}, {Xie}, \& {IXPE Collaboration}}]{2024ApJ...968...76I}
{Ingram}, A., {Bollemeijer}, N., {Veledina}, A., {et~al.} 2024, \apj, 968, 76, \dodoi{10.3847/1538-4357/ad3faf}

\bibitem[{{Iyer} {et~al.}(2023){Iyer}, {Kiss}, {Pearce}, {Stana}, {Awaki}, {Bose}, {Dasgupta}, {De Geronimo}, {Gau}, {Hakamata}, {Ishida}, {Ishiwata}, {Kamogawa}, {Kislat}, {Kitaguchi}, {Krawczynski}, {Lisalda}, {Maeda}, {Matsumoto}, {Miyamoto}, {Miyazawa}, {Mizuno}, {Rauch}, {Cavero}, {Sakamoto}, {Sato}, {Spooner}, {Takahashi}, {Takeo}, {Tamagawa}, {Uchida}, {West}, {Wimalasena}, \& {Yoshimoto}}]{2023NIMPA104867975I}
{Iyer}, N.~K., {Kiss}, M., {Pearce}, M., {et~al.} 2023, Nuclear Instruments and Methods in Physics Research A, 1048, 167975, \dodoi{10.1016/j.nima.2022.167975}

\bibitem[{{Johnston} {et~al.}(1982){Johnston}, {Elvis}, {Kjer}, \& {Shen}}]{1982ApJ...262...61J}
{Johnston}, K.~J., {Elvis}, M., {Kjer}, D., \& {Shen}, B.~S.~P. 1982, \apj, 262, 61, \dodoi{10.1086/160396}

\bibitem[{{Jourdain} {et~al.}(2012){Jourdain}, {Roques}, {Chauvin}, \& {Clark}}]{2012ApJ...761...27J}
{Jourdain}, E., {Roques}, J.~P., {Chauvin}, M., \& {Clark}, D.~J. 2012, \apj, 761, 27, \dodoi{10.1088/0004-637X/761/1/27}

\bibitem[{{Kantzas} {et~al.}(2021){Kantzas}, {Markoff}, {Beuchert}, {Lucchini}, {Chhotray}, {Ceccobello}, {Tetarenko}, {Miller-Jones}, {Bremer}, {Garcia}, {Grinberg}, {Uttley}, \& {Wilms}}]{2021MNRAS.500.2112K}
{Kantzas}, D., {Markoff}, S., {Beuchert}, T., {et~al.} 2021, \mnras, 500, 2112, \dodoi{10.1093/mnras/staa3349}

\bibitem[{{Kislat} {et~al.}(2015){Kislat}, {Clark}, {Beilicke}, \& {Krawczynski}}]{2015APh....68...45K}
{Kislat}, F., {Clark}, B., {Beilicke}, M., \& {Krawczynski}, H. 2015, Astroparticle Physics, 68, 45, \dodoi{10.1016/j.astropartphys.2015.02.007}

\bibitem[{Kislat \& Spooner(2024)}]{Kislat2024}
Kislat, F., \& Spooner, S. 2024, in Handbook of X-ray and Gamma-ray Astrophysics, ed. C.~Bambi \& A.~Santangelo (Singapore: Springer Nature Singapore), 5853, \dodoi{10.1007/978-981-16-4544-0_141-1}

\bibitem[{Kiss \& Pearce(2024)}]{Kiss2024}
Kiss, M., \& Pearce, M. 2024, in Handbook of X-ray and Gamma-ray Astrophysics, ed. C.~Bambi \& A.~Santangelo (Singapore: Springer Nature Singapore), 5683, \dodoi{10.1007/978-981-16-4544-0_141-1}

\bibitem[{{Krawczynski} \& {Beheshtipour}(2022)}]{2022ApJ...934....4K}
{Krawczynski}, H., \& {Beheshtipour}, B. 2022, \apj, 934, 4, \dodoi{10.3847/1538-4357/ac7725}

\bibitem[{{Krawczynski} {et~al.}(2022){Krawczynski}, {Muleri}, {Dov{\v{c}}iak}, {Veledina}, {Rodriguez Cavero}, {Svoboda}, {Ingram}, {Matt}, {Garcia}, {Loktev}, {Negro}, {Poutanen}, {Kitaguchi}, {Podgorn{\'y}}, {Rankin}, {Zhang}, {Berdyugin}, {Berdyugina}, {Bianchi}, {Blinov}, {Capitanio}, {Di Lalla}, {Draghis}, {Fabiani}, {Kagitani}, {Kravtsov}, {Kiehlmann}, {Latronico}, {Lutovinov}, {Mandarakas}, {Marin}, {Marinucci}, {Miller}, {Mizuno}, {Molkov}, {Omodei}, {Petrucci}, {Ratheesh}, {Sakanoi}, {Semena}, {Skalidis}, {Soffitta}, {Tennant}, {Thalhammer}, {Tombesi}, {Weisskopf}, {Wilms}, {Zhang}, {Agudo}, {Antonelli}, {Bachetti}, {Baldini}, {Baumgartner}, {Bellazzini}, {Bongiorno}, {Bonino}, {Brez}, {Bucciantini}, {Castellano}, {Cavazzuti}, {Ciprini}, {Costa}, {De Rosa}, {Del Monte}, {Di Gesu}, {Di Marco}, {Donnarumma}, {Doroshenko}, {Ehlert}, {Enoto}, {Evangelista}, {Ferrazzoli}, {Gunji}, {Hayashida}, {Heyl}, {Iwakiri}, {Jorstad}, {Karas}, {Kolodziejczak}, {La Monaca}, {Liodakis}, {Maldera}, {Manfreda},
  {Marscher}, {Marshall}, {Mitsuishi}, {Ng}, {O{\textquoteright}Dell}, {Oppedisano}, {Papitto}, {Pavlov}, {Peirson}, {Perri}, {Pesce-Rollins}, {Pilia}, {Possenti}, {Puccetti}, {Ramsey}, {Romani}, {Sgr{\`o}}, {Slane}, {Spandre}, {Tamagawa}, {Tavecchio}, {Taverna}, {Tawara}, {Thomas}, {Trois}, {Tsygankov}, {Turolla}, {Vink}, {Wu}, {Xie}, \& {Zane}}]{2022Sci...378..650K}
{Krawczynski}, H., {Muleri}, F., {Dov{\v{c}}iak}, M., {et~al.} 2022, Science, 378, 650, \dodoi{10.1126/science.add5399}

\bibitem[{{Krimm} {et~al.}(2013){Krimm}, {Holland}, {Corbet}, {Pearlman}, {Romano}, {Kennea}, {Bloom}, {Barthelmy}, {Baumgartner}, {Cummings}, {Gehrels}, {Lien}, {Markwardt}, {Palmer}, {Sakamoto}, {Stamatikos}, \& {Ukwatta}}]{2013ApJS..209...14K}
{Krimm}, H.~A., {Holland}, S.~T., {Corbet}, R.~H.~D., {et~al.} 2013, \apjs, 209, 14, \dodoi{10.1088/0067-0049/209/1/14}

\bibitem[{{Kuramoto} {et~al.}(2023){Kuramoto}, {Matsumoto}, {Awaki}, {Bose}, {Braun}, {Chun}, {De Geronimo}, {Eric A.}, {Errando}, {Fukazawa}, {Furusawa}, {Gadson}, {Gau}, {Guarino}, {Gunji}, {Harmon}, {Heatwole}, {Hossen}, {Ishibashi}, {Ishida}, {Iyer}, {Kamogawa}, {Kislat}, {Kiss}, {Krawczynski}, {Lanzi}, {Lisalda}, {Maeda}, {Miyamoto}, {Miyazawa}, {Okajima}, {Pearce}, {Peterson}, {Rauch}, {Rodriguez Cavero}, {Ryde}, {Simburger}, {Spooner}, {Stana}, {Stuchlik}, {Takahashi}, {Takeo}, {Tamagawa}, {Uchida}, \& {West}}]{2023SPIE12679E..1BK}
{Kuramoto}, H., {Matsumoto}, H., {Awaki}, H., {et~al.} 2023, in Society of Photo-Optical Instrumentation Engineers (SPIE) Conference Series, Vol. 12679, Optics for EUV, X-Ray, and Gamma-Ray Astronomy XI, ed. S.~L. {O'Dell}, J.~A. {Gaskin}, G.~{Pareschi}, \& D.~{Spiga}, 126791B, \dodoi{10.1117/12.2676572}

\bibitem[{{Laurent} {et~al.}(2011){Laurent}, {Rodriguez}, {Wilms}, {Cadolle Bel}, {Pottschmidt}, \& {Grinberg}}]{2011Sci...332..438L}
{Laurent}, P., {Rodriguez}, J., {Wilms}, J., {et~al.} 2011, Science, 332, 438, \dodoi{10.1126/science.1200848}

\bibitem[{{Li} {et~al.}(2025){Li}, {Zhao}, {Feng}, {Tao}, \& {Tsygankov}}]{2025arXiv250404775L}
{Li}, H., {Zhao}, Q.-C., {Feng}, H., {Tao}, L., \& {Tsygankov}, S.~S. 2025, arXiv e-prints, arXiv:2504.04775, \dodoi{10.48550/arXiv.2504.04775}

\bibitem[{{Li} {et~al.}(2009){Li}, {Narayan}, \& {McClintock}}]{2009ApJ...691..847L}
{Li}, L.-X., {Narayan}, R., \& {McClintock}, J.~E. 2009, \apj, 691, 847, \dodoi{10.1088/0004-637X/691/1/847}

\bibitem[{{Liodakis} {et~al.}(2022){Liodakis}, {Marscher}, {Agudo}, {Berdyugin}, {Bernardos}, {Bonnoli}, {Borman}, {Casadio}, {Casanova}, {Cavazzuti}, {Rodriguez Cavero}, {Di Gesu}, {Di Lalla}, {Donnarumma}, {Ehlert}, {Errando}, {Escudero}, {Garc{\'\i}a-Comas}, {Ag{\'\i}s-Gonz{\'a}lez}, {Husillos}, {Jormanainen}, {Jorstad}, {Kagitani}, {Kopatskaya}, {Kravtsov}, {Krawczynski}, {Lindfors}, {Larionova}, {Madejski}, {Marin}, {Marchini}, {Marshall}, {Morozova}, {Massaro}, {Masiero}, {Mawet}, {Middei}, {Millar-Blanchaer}, {Myserlis}, {Negro}, {Nilsson}, {O'Dell}, {Omodei}, {Pacciani}, {Paggi}, {Panopoulou}, {Peirson}, {Perri}, {Petrucci}, {Poutanen}, {Puccetti}, {Romani}, {Sakanoi}, {Savchenko}, {Sota}, {Tavecchio}, {Tinyanont}, {Vasilyev}, {Weaver}, {Zhovtan}, {Antonelli}, {Bachetti}, {Baldini}, {Baumgartner}, {Bellazzini}, {Bianchi}, {Bongiorno}, {Bonino}, {Brez}, {Bucciantini}, {Capitanio}, {Castellano}, {Ciprini}, {Costa}, {De Rosa}, {Del Monte}, {Di Marco}, {Doroshenko}, {Dov{\v{c}}iak}, {Enoto},
  {Evangelista}, {Fabiani}, {Ferrazzoli}, {Garcia}, {Gunji}, {Hayashida}, {Heyl}, {Iwakiri}, {Karas}, {Kitaguchi}, {Kolodziejczak}, {La Monaca}, {Latronico}, {Maldera}, {Manfreda}, {Marinucci}, {Matt}, {Mitsuishi}, {Mizuno}, {Muleri}, {Ng}, {Oppedisano}, {Papitto}, {Pavlov}, {Pesce-Rollins}, {Pilia}, {Possenti}, {Ramsey}, {Rankin}, {Ratheesh}, {Sgr{\'o}}, {Slane}, {Soffitta}, {Spandre}, {Tamagawa}, {Taverna}, {Tawara}, {Tennant}, {Thomas}, {Tombesi}, {Trois}, {Tsygankov}, {Turolla}, {Vink}, {Weisskopf}, {Wu}, {Xie}, \& {Zane}}]{2022Natur.611..677L}
{Liodakis}, I., {Marscher}, A.~P., {Agudo}, I., {et~al.} 2022, \nat, 611, 677, \dodoi{10.1038/s41586-022-05338-0}

\bibitem[{{Long} {et~al.}(1980){Long}, {Chanan}, \& {Novick}}]{1980ApJ...238..710L}
{Long}, K.~S., {Chanan}, G.~A., \& {Novick}, R. 1980, \apj, 238, 710, \dodoi{10.1086/158027}

\bibitem[{{Madsen} {et~al.}(2015){Madsen}, {Reynolds}, {Harrison}, {An}, {Boggs}, {Christensen}, {Craig}, {Fryer}, {Grefenstette}, {Hailey}, {Markwardt}, {Nynka}, {Stern}, {Zoglauer}, \& {Zhang}}]{2015ApJ...801...66M}
{Madsen}, K.~K., {Reynolds}, S., {Harrison}, F., {et~al.} 2015, \apj, 801, 66, \dodoi{10.1088/0004-637X/801/1/66}

\bibitem[{{Mastroserio} {et~al.}(2024){Mastroserio}, {De Marco}, {Baglio}, {Carotenuto}, {Fabiani}, {Russell}, {Capitanio}, {Cavecchi}, {Motta}, {Russell}, {Dovciak}, {Del Santo}, {Alabarta}, {Ambrifi}, {Campana}, {Casella}, {Covino}, {Illiano}, {Kara}, {Lai}, {Lodato}, {Manca}, {Mariani}, {Marino}, {Miceli}, {Saikia}, {Shaw}, {Svoboda}, {Vincentelli}, \& {Wang}}]{2024arXiv240806856M}
{Mastroserio}, G., {De Marco}, B., {Baglio}, M.~C., {et~al.} 2024, arXiv e-prints, arXiv:2408.06856, \dodoi{10.48550/arXiv.2408.06856}

\bibitem[{{Miller-Jones} {et~al.}(2021){Miller-Jones}, {Bahramian}, {Orosz}, {Mandel}, {Gou}, {Maccarone}, {Neijssel}, {Zhao}, {Zi{\'o}{\l}kowski}, {Reid}, {Uttley}, {Zheng}, {Byun}, {Dodson}, {Grinberg}, {Jung}, {Kim}, {Marcote}, {Markoff}, {Rioja}, {Rushton}, {Russell}, {Sivakoff}, {Tetarenko}, {Tudose}, \& {Wilms}}]{2021Sci...371.1046M}
{Miller-Jones}, J. C.~A., {Bahramian}, A., {Orosz}, J.~A., {et~al.} 2021, Science, 371, 1046, \dodoi{10.1126/science.abb3363}

\bibitem[{{Moscibrodzka}(2024)}]{2024Ap&SS.369...68M}
{Moscibrodzka}, M. 2024, \apss, 369, 68, \dodoi{10.1007/s10509-024-04333-3}

\bibitem[{{Poutanen} {et~al.}(2023){Poutanen}, {Veledina}, \& {Beloborodov}}]{2023ApJ...949L..10P}
{Poutanen}, J., {Veledina}, A., \& {Beloborodov}, A.~M. 2023, \apjl, 949, L10, \dodoi{10.3847/2041-8213/acd33e}

\bibitem[{{Poutanen} \& {Vilhu}(1993)}]{1993A&A...275..337P}
{Poutanen}, J., \& {Vilhu}, O. 1993, \aap, 275, 337

\bibitem[{{Quinn}(2012)}]{2012A&A...538A..65Q}
{Quinn}, J.~L. 2012, \aap, 538, A65, \dodoi{10.1051/0004-6361/201015785}

\bibitem[{{Rodriguez} {et~al.}(2015){Rodriguez}, {Grinberg}, {Laurent}, {Cadolle Bel}, {Pottschmidt}, {Pooley}, {Bodaghee}, {Wilms}, \& {Gouiff{\`e}s}}]{2015ApJ...807...17R}
{Rodriguez}, J., {Grinberg}, V., {Laurent}, P., {et~al.} 2015, \apj, 807, 17, \dodoi{10.1088/0004-637X/807/1/17}

\bibitem[{{Saade} {et~al.}(2024){Saade}, {Kaaret}, {Liodakis}, \& {Ehlert}}]{2024ApJ...974..101S}
{Saade}, M.~L., {Kaaret}, P., {Liodakis}, I., \& {Ehlert}, S.~R. 2024, \apj, 974, 101, \dodoi{10.3847/1538-4357/ad73a3}

\bibitem[{{Schnittman} \& {Krolik}(2009)}]{2009ApJ...701.1175S}
{Schnittman}, J.~D., \& {Krolik}, J.~H. 2009, \apj, 701, 1175, \dodoi{10.1088/0004-637X/701/2/1175}

\bibitem[{{Schnittman} \& {Krolik}(2010)}]{2010ApJ...712..908S}
---. 2010, \apj, 712, 908, \dodoi{10.1088/0004-637X/712/2/908}

\bibitem[{{Sridhar} {et~al.}(2025){Sridhar}, {Ripperda}, {Sironi}, {Davelaar}, \& {Beloborodov}}]{2025ApJ...979..199S}
{Sridhar}, N., {Ripperda}, B., {Sironi}, L., {Davelaar}, J., \& {Beloborodov}, A.~M. 2025, \apj, 979, 199, \dodoi{10.3847/1538-4357/ada385}

\bibitem[{{Steiner} {et~al.}(2024){Steiner}, {Nathan}, {Hu}, {Krawczynski}, {Dov{\v{c}}iak}, {Veledina}, {Muleri}, {Svoboda}, {Alabarta}, {Parra}, {Bhargava}, {Matt}, {Poutanen}, {Petrucci}, {Tennant}, {Baglio}, {Baldini}, {Barnier}, {Bhattacharyya}, {Bianchi}, {Brigitte}, {Cabezas}, {Cangemi}, {Capitanio}, {Casey}, {Rodriguez Cavero}, {Castellano}, {Cavazzuti}, {Chun}, {Churazov}, {Costa}, {Di Lalla}, {Di Marco}, {Egron}, {Ewing}, {Fabiani}, {Garc{\'\i}a}, {Green}, {Grinberg}, {Hadrava}, {Ingram}, {Kaaret}, {Kislat}, {Kitaguchi}, {Kravtsov}, {Kub{\'a}tov{\'a}}, {La Monaca}, {Latronico}, {Loktev}, {Malacaria}, {Marin}, {Marinucci}, {Maryeva}, {Mastroserio}, {Mizuno}, {Negro}, {Omodei}, {Podgorn{\'y}}, {Rankin}, {Ratheesh}, {Rhodes}, {Russell}, {{\v{S}}lechta}, {Soffitta}, {Spooner}, {Suleimanov}, {Tombesi}, {Trushkin}, {Weisskopf}, {Zane}, {Zdziarski}, {Zhang}, {Zhang}, {Zhou}, {Agudo}, {Antonelli}, {Bachetti}, {Baumgartner}, {Bellazzini}, {Bongiorno}, {Bonino}, {Brez}, {Bucciantini}, {Chen}, {Ciprini}, {De
  Rosa}, {Del Monte}, {Di Gesu}, {Donnarumma}, {Doroshenko}, {Ehlert}, {Enoto}, {Evangelista}, {Ferrazzoli}, {Gunji}, {Hayashida}, {Heyl}, {Iwakiri}, {Jorstad}, {Karas}, {Kolodziejczak}, {Liodakis}, {Maldera}, {Manfreda}, {Marscher}, {Marshall}, {Massaro}, {Mitsuishi}, {Ng}, {O'Dell}, {Oppedisano}, {Papitto}, {Pavlov}, {Peirson}, {Perri}, {Pesce-Rollins}, {Pilia}, {Possenti}, {Puccetti}, {Ramsey}, {Roberts}, {Romani}, {Sgr{\`o}}, {Slane}, {Spandre}, {Swartz}, {Tamagawa}, {Tavecchio}, {Taverna}, {Tawara}, {Thomas}, {Trois}, {Tsygankov}, {Turolla}, {Vink}, {Wu}, \& {Xie}}]{2024ApJ...969L..30S}
{Steiner}, J.~F., {Nathan}, E., {Hu}, K., {et~al.} 2024, \apjl, 969, L30, \dodoi{10.3847/2041-8213/ad58e4}

\bibitem[{Stuchlik(2017)}]{doi:10.2514/6.2017-3609}
Stuchlik, D. 2017, The NASA Wallops Arc-Second Pointer (WASP) System for Precision Pointing of Scientific Balloon Instruments and Telescopes, \dodoi{10.2514/6.2017-3609}

\bibitem[{{Sunyaev} \& {Titarchuk}(1985)}]{1985A&A...143..374S}
{Sunyaev}, R.~A., \& {Titarchuk}, L.~G. 1985, \aap, 143, 374

\bibitem[{{Taverna} {et~al.}(2022){Taverna}, {Turolla}, {Muleri}, {Heyl}, {Zane}, {Baldini}, {Gonz{\'a}lez-Caniulef}, {Bachetti}, {Rankin}, {Caiazzo}, {Di Lalla}, {Doroshenko}, {Errando}, {Gau}, {K{\i}rm{\i}z{\i}bayrak}, {Krawczynski}, {Negro}, {Ng}, {Omodei}, {Possenti}, {Tamagawa}, {Uchiyama}, {Weisskopf}, {Agudo}, {Antonelli}, {Baumgartner}, {Bellazzini}, {Bianchi}, {Bongiorno}, {Bonino}, {Brez}, {Bucciantini}, {Capitanio}, {Castellano}, {Cavazzuti}, {Ciprini}, {Costa}, {De Rosa}, {Del Monte}, {Di Gesu}, {Di Marco}, {Donnarumma}, {Dov{\v{c}}iak}, {Ehlert}, {Enoto}, {Evangelista}, {Fabiani}, {Ferrazzoli}, {Garcia}, {Gunji}, {Hayashida}, {Iwakiri}, {Jorstad}, {Karas}, {Kitaguchi}, {Kolodziejczak}, {La Monaca}, {Latronico}, {Liodakis}, {Maldera}, {Manfreda}, {Marin}, {Marinucci}, {Marscher}, {Marshall}, {Matt}, {Mitsuishi}, {Mizuno}, {Ng}, {O{\textquoteright}Dell}, {Oppedisano}, {Papitto}, {Pavlov}, {Peirson}, {Perri}, {Pesce-Rollins}, {Pilia}, {Poutanen}, {Puccetti}, {Ramsey}, {Ratheesh}, {Romani},
  {Sgr{\`o}}, {Slane}, {Soffitta}, {Spandre}, {Tavecchio}, {Tawara}, {Tennant}, {Thomas}, {Tombesi}, {Trois}, {Tsygankov}, {Vink}, {Wu}, \& {Xie}}]{2022Sci...378..646T}
{Taverna}, R., {Turolla}, R., {Muleri}, F., {et~al.} 2022, Science, 378, 646, \dodoi{10.1126/science.add0080}

\bibitem[{{Tomsick} {et~al.}(2024){Tomsick}, {Boggs}, {Zoglauer}, {Hartmann}, {Ajello}, {Burns}, {Fryer}, {Karwin}, {Kierans}, {Lowell}, {Malzac}, {Roberts}, {Saint-Hilaire}, {Shih}, {Siegert}, {Sleator}, {Takahashi}, {Tavecchio}, {Wulf}, {Beechert}, {Gulick}, {Joens}, {Lazar}, {Neights}, {Martinez Oliveros}, {Matsumoto}, {Melia}, {Yoneda}, {Amman}, {Bal}, {von Ballmoos}, {Bates}, {B{\"o}ttcher}, {Bulgarelli}, {Cavazzuti}, {Chang}, {Chen}, {Chu}, {Ciabattoni}, {Costamante}, {Dreyer}, {Fioretti}, {Fenu}, {Gallego}, {Ghirlanda}, {Grove}, {Huang}, {Jean}, {Khatiya}, {Kn{\"o}dlseder}, {Kraus}, {Leising}, {Lewis}, {Lommler}, {Marcotulli}, {Martinez Castellanos}, {Mittal}, {Negro}, {Al Nussirat}, {Nakazawa}, {Oberlack}, {Palmore}, {Panebianco}, {Parmiggiani}, {Pike}, {Rogers}, {Schutte}, {Sheng}, {Smale}, {Smith}, {Trigg}, {Venters}, {Watanabe}, \& {Zhang}}]{2024icrc.confE.745T}
{Tomsick}, J., {Boggs}, S., {Zoglauer}, A., {et~al.} 2024, in 38th International Cosmic Ray Conference, 745, \dodoi{10.48550/arXiv.2308.12362}

\bibitem[{{Tomsick} {et~al.}(2022){Tomsick}, {Lowell}, {Lazar}, {Sleator}, \& {Zoglauer}}]{2022hxga.book...73T}
{Tomsick}, J.~A., {Lowell}, A., {Lazar}, H., {Sleator}, C., \& {Zoglauer}, A. 2022, in Handbook of X-ray and Gamma-ray Astrophysics, ed. C.~{Bambi} \& A.~{Sangangelo}, 73, \dodoi{10.1007/978-981-16-4544-0_145-1}

\bibitem[{{Ulvestad} {et~al.}(1998){Ulvestad}, {Roy}, {Colbert}, \& {Wilson}}]{1998ApJ...496..196U}
{Ulvestad}, J.~S., {Roy}, A.~L., {Colbert}, E. J.~M., \& {Wilson}, A.~S. 1998, \apj, 496, 196, \dodoi{10.1086/305382}

\bibitem[{{Unger} {et~al.}(1987){Unger}, {Lawrence}, {Wilson}, {Elvis}, \& {Wright}}]{1987MNRAS.228..521U}
{Unger}, S.~W., {Lawrence}, A., {Wilson}, A.~S., {Elvis}, M., \& {Wright}, A.~E. 1987, \mnras, 228, 521, \dodoi{10.1093/mnras/228.3.521}

\bibitem[{{Veledina} {et~al.}(2023){Veledina}, {Muleri}, {Dov{\v{c}}iak}, {Poutanen}, {Ratheesh}, {Capitanio}, {Matt}, {Soffitta}, {Tennant}, {Negro}, {Kaaret}, {Costa}, {Ingram}, {Svoboda}, {Krawczynski}, {Bianchi}, {Steiner}, {Garc{\'\i}a}, {Kravtsov}, {Nitindala}, {Ewing}, {Mastroserio}, {Marinucci}, {Ursini}, {Tombesi}, {Tsygankov}, {Yang}, {Weisskopf}, {Trushkin}, {Egron}, {Iacolina}, {Pilia}, {Marra}, {Miku{\v{s}}incov{\'a}}, {Nathan}, {Parra}, {Petrucci}, {Podgorn{\'y}}, {Tugliani}, {Zane}, {Zhang}, {Agudo}, {Antonelli}, {Bachetti}, {Baldini}, {Baumgartner}, {Bellazzini}, {Bongiorno}, {Bonino}, {Brez}, {Bucciantini}, {Castellano}, {Cavazzuti}, {Chen}, {Ciprini}, {De Rosa}, {Del Monte}, {Di Gesu}, {Di Lalla}, {Di Marco}, {Donnarumma}, {Doroshenko}, {Ehlert}, {Enoto}, {Evangelista}, {Fabiani}, {Ferrazzoli}, {Gunji}, {Hayashida}, {Heyl}, {Iwakiri}, {Jorstad}, {Karas}, {Kislat}, {Kitaguchi}, {Kolodziejczak}, {La Monaca}, {Latronico}, {Liodakis}, {Maldera}, {Manfreda}, {Marin}, {Marscher}, {Marshall},
  {Massaro}, {Mitsuishi}, {Mizuno}, {Ng}, {O'Dell}, {Omodei}, {Oppedisano}, {Papitto}, {Pavlov}, {Peirson}, {Perri}, {Pesce-Rollins}, {Possenti}, {Puccetti}, {Ramsey}, {Rankin}, {Roberts}, {Romani}, {Sgr{\`o}}, {Slane}, {Spandre}, {Swartz}, {Tamagawa}, {Tavecchio}, {Taverna}, {Tawara}, {Thomas}, {Trois}, {Turolla}, {Vink}, {Wu}, \& {Xie}}]{2023ApJ...958L..16V}
{Veledina}, A., {Muleri}, F., {Dov{\v{c}}iak}, M., {et~al.} 2023, \apjl, 958, L16, \dodoi{10.3847/2041-8213/ad0781}

\bibitem[{{Vink} {et~al.}(2022){Vink}, {Prokhorov}, {Ferrazzoli}, {Slane}, {Zhou}, {Asakura}, {Baldini}, {Bucciantini}, {Costa}, {Di Marco}, {Heyl}, {Marin}, {Mizuno}, {Ng}, {Pesce-Rollins}, {Ramsey}, {Rankin}, {Ratheesh}, {Sgr{\'o}}, {Soffitta}, {Swartz}, {Tamagawa}, {Weisskopf}, {Yang}, {Bellazzini}, {Bonino}, {Cavazzuti}, {Costamante}, {Di Lalla}, {Latronico}, {Maldera}, {Manfreda}, {Massaro}, {Mitsuishi}, {Omodei}, {Oppedisano}, {Zane}, {Agudo}, {Antonelli}, {Bachetti}, {Baumgartner}, {Bianchi}, {Bongiorno}, {Brez}, {Capitanio}, {Castellano}, {Ciprini}, {De Rosa}, {Del Monte}, {Di Gesu}, {Donnarumma}, {Doroshenko}, {Dov{\v{c}}iak}, {Ehlert}, {Enoto}, {Evangelista}, {Fabiani}, {Garcia}, {Gunji}, {Hayashida}, {Iwakiri}, {Jorstad}, {Karas}, {Kitaguchi}, {Kolodziejczak}, {Krawczynski}, {La Monaca}, {Liodakis}, {Marinucci}, {Marscher}, {Marshall}, {Matt}, {Muleri}, {O'Dell}, {Papitto}, {Pavlov}, {Peirson}, {Perri}, {Pilia}, {Possenti}, {Poutanen}, {Puccetti}, {Romani}, {Spandre}, {Tavecchio}, {Taverna},
  {Tawara}, {Tennant}, {Thomas}, {Tombesi}, {Trois}, {Tsygankov}, {Turolla}, {Wu}, \& {Xie}}]{2022ApJ...938...40V}
{Vink}, J., {Prokhorov}, D., {Ferrazzoli}, R., {et~al.} 2022, \apj, 938, 40, \dodoi{10.3847/1538-4357/ac8b7b}

\bibitem[{{Walton} {et~al.}(2016){Walton}, {Tomsick}, {Madsen}, {Grinberg}, {Barret}, {Boggs}, {Christensen}, {Clavel}, {Craig}, {Fabian}, {Fuerst}, {Hailey}, {Harrison}, {Miller}, {Parker}, {Rahoui}, {Stern}, {Tao}, {Wilms}, \& {Zhang}}]{2016ApJ...826...87W}
{Walton}, D.~J., {Tomsick}, J.~A., {Madsen}, K.~K., {et~al.} 2016, \apj, 826, 87, \dodoi{10.3847/0004-637X/826/1/87}

\bibitem[{{Weisskopf} {et~al.}(2022){Weisskopf}, {Soffitta}, {Baldini}, {Ramsey}, {O'Dell}, {Romani}, {Matt}, {Deininger}, {Baumgartner}, {Bellazzini}, {Costa}, {Kolodziejczak}, {Latronico}, {Marshall}, {Muleri}, {Bongiorno}, {Tennant}, {Bucciantini}, {Dovciak}, {Marin}, {Marscher}, {Poutanen}, {Slane}, {Turolla}, {Kalinowski}, {Di Marco}, {Fabiani}, {Minuti}, {La Monaca}, {Pinchera}, {Rankin}, {Sgro'}, {Trois}, {Xie}, {Alexander}, {Allen}, {Amici}, {Andersen}, {Antonelli}, {Antoniak}, {Attina'}, {Barbanera}, {Bachetti}, {Baggett}, {Bladt}, {Brez}, {Bonino}, {Boree}, {Borotto}, {Breeding}, {Brienza}, {Bygott}, {Caporale}, {Cardelli}, {Carpentiero}, {Castellano}, {Castronuovo}, {Cavalli}, {Cavazzuti}, {Ceccanti}, {Centrone}, {Citraro}, {D'Amico}, {D'Alba}, {Di Gesu}, {Del Monte}, {Dietz}, {Di Lalla}, {Di Persio}, {Dolan}, {Donnarumma}, {Evangelista}, {Ferrant}, {Ferrazzoli}, {Ferrie}, {Footdale}, {Forsyth}, {Foster}, {Garelick}, {Gunji}, {Gurnee}, {Head}, {Hibbard}, {Johnson}, {Kelly}, {Kilaru}, {Lefevre}, {Le
  Roy}, {Loffredo}, {Lorenzi}, {Lucchesi}, {Maddox}, {Magazzu}, {Maldera}, {Manfreda}, {Mangraviti}, {Marengo}, {Marrocchesi}, {Massaro}, {Mauger}, {McCracken}, {McEachen}, {Mize}, {Mereu}, {Mitchell}, {Mitsuishi}, {Morbidini}, {Mosti}, {Nasimi}, {Negri}, {Negro}, {Nguyen}, {Nitschke}, {Nuti}, {Onizuka}, {Oppedisano}, {Orsini}, {Osborne}, {Pacheco}, {Paggi}, {Painter}, {Pavelitz}, {Pentz}, {Piazzolla}, {Perri}, {Pesce-Rollins}, {Peterson}, {Pilia}, {Profeti}, {Puccetti}, {Ranganathan}, {Ratheesh}, {Reedy}, {Root}, {Rubini}, {Ruswick}, {Sanchez}, {Sarra}, {Santoli}, {Scalise}, {Sciortino}, {Schroeder}, {Seek}, {Sosdian}, {Spandre}, {Speegle}, {Tamagawa}, {Tardiola}, {Tobia}, {Thomas}, {Valerie}, {Vimercati}, {Walden}, {Weddendorf}, {Wedmore}, {Welch}, {Zanetti}, \& {Zanetti}}]{ixpe}
{Weisskopf}, M.~C., {Soffitta}, P., {Baldini}, L., {et~al.} 2022, Journal of Astronomical Telescopes, Instruments, and Systems, 8, 026002, \dodoi{10.1117/1.JATIS.8.2.026002}

\bibitem[{{Wilms} {et~al.}(2006){Wilms}, {Nowak}, {Pottschmidt}, {Pooley}, \& {Fritz}}]{2006A&A...447..245W}
{Wilms}, J., {Nowak}, M.~A., {Pottschmidt}, K., {Pooley}, G.~G., \& {Fritz}, S. 2006, \aap, 447, 245, \dodoi{10.1051/0004-6361:20053938}

\bibitem[{{Wood} {et~al.}(2024){Wood}, {Miller-Jones}, {Bahramian}, {Tingay}, {Prabu}, {Russell}, {Atri}, {Carotenuto}, {Altamirano}, {Motta}, {Hyland}, {Reynolds}, {Weston}, {Fender}, {K{\"o}rding}, {Maitra}, {Markoff}, {Migliari}, {Russell}, {Sarazin}, {Sivakoff}, {Soria}, {Tetarenko}, \& {Tudose}}]{2024ApJ...971L...9W}
{Wood}, C.~M., {Miller-Jones}, J. C.~A., {Bahramian}, A., {et~al.} 2024, \apjl, 971, L9, \dodoi{10.3847/2041-8213/ad6572}

\bibitem[{{Xie} {et~al.}(2022){Xie}, {Di Marco}, {La Monaca}, {Liu}, {Muleri}, {Bucciantini}, {Romani}, {Costa}, {Rankin}, {Soffitta}, {Bachetti}, {Di Lalla}, {Fabiani}, {Ferrazzoli}, {Gunji}, {Latronico}, {Negro}, {Omodei}, {Pilia}, {Trois}, {Watanabe}, {Agudo}, {Antonelli}, {Baldini}, {Baumgartner}, {Bellazzini}, {Bianchi}, {Bongiorno}, {Bonino}, {Brez}, {Capitanio}, {Castellano}, {Cavazzuti}, {Ciprini}, {De Rosa}, {Del Monte}, {Di Gesu}, {Donnarumma}, {Doroshenko}, {Dov{\v{c}}iak}, {Ehlert}, {Enoto}, {Evangelista}, {Garcia}, {Hayashida}, {Heyl}, {Iwakiri}, {Jorstad}, {Karas}, {Kitaguchi}, {Kolodziejczak}, {Krawczynski}, {Liodakis}, {Maldera}, {Manfreda}, {Marin}, {Marinucci}, {Marscher}, {Marshall}, {Massaro}, {Matt}, {Mitsuishi}, {Mizuno}, {Ng}, {O'Dell}, {Oppedisano}, {Papitto}, {Pavlov}, {Peirson}, {Perri}, {Pesce-Rollins}, {Petrucci}, {Possenti}, {Poutanen}, {Puccetti}, {Ramsey}, {Ratheesh}, {Sgr{\'o}}, {Slane}, {Spandre}, {Tamagawa}, {Tavecchio}, {Taverna}, {Tawara}, {Tennant}, {Thomas}, {Tombesi},
  {Tsygankov}, {Turolla}, {Vink}, {Weisskopf}, {Wu}, \& {Zane}}]{2022Natur.612..658X}
{Xie}, F., {Di Marco}, A., {La Monaca}, F., {et~al.} 2022, \nat, 612, 658, \dodoi{10.1038/s41586-022-05476-5}

\bibitem[{{Zhang} {et~al.}(2019){Zhang}, {Dov{\v{c}}iak}, \& {Bursa}}]{2019ApJ...875..148Z}
{Zhang}, W., {Dov{\v{c}}iak}, M., \& {Bursa}, M. 2019, \apj, 875, 148, \dodoi{10.3847/1538-4357/ab1261}

\end{thebibliography}


\label{lastpage}
\end{document}